 \documentclass[12pt,preprint]{aastex}

\usepackage{epsfig}

\def\green{f_{_{\rm G}}}
\def\ngreen{n_{_{\rm G}}}
\def\ugreen{U_{_{\rm G}}}

\newcommand{\vff}{v_{\rm ff}}

\newcommand{\gapprox}{\lower.4ex\hbox{$\;\buildrel >\over{\scriptstyle\sim}\;$}}
\newcommand{\lapprox}{\lower.4ex\hbox{$\;\buildrel <\over{\scriptstyle\sim}\;$}}
\newcommand{\begeq}{\begin{equation}}
\newcommand{\fineq}{\end{equation}}
\newcommand{\msun}{M_\odot} 
\newcommand{\Msun}{M_\odot} 
\newcommand{\sig}{\sigma_{_{\rm T}}}
\newcommand{\xlum}{L_{\rm X}}

\def\ellprime0{\ell'_0}

\def\sig{\sigma_{_{\rm T}}}

\def\sigpar{\sigma_\|}
\def\sigperp{\sigma_\perp}
\def\taupar{\tau_\|}
\def\tauperp{\tau_\perp}

\def\colrad{r_0}
\def\starad{R_*}
\def\starmass{M_*}

\def\greenphoton{\dot N_\epsilon^{\rm G}}
\def\photonparticular{\dot N_\epsilon}
\def\greencolumn{\Phi_\epsilon^{\rm G}}
\def\column{\Phi_\epsilon}

\slugcomment{accepted by \apj}

\shorttitle{X-Ray Pulsar Spectral Formation}
\shortauthors{Becker \& Wolff}

\begin{document}

\title{SPECTRAL FORMATION IN X-RAY PULSARS: \break
BULK COMPTONIZATION IN THE ACCRETION SHOCK}

\author{Peter A. Becker\altaffilmark{1}$^,$\altaffilmark{2}}

\affil{Center for Earth Observing and Space Research, \break
George Mason University \break
Fairfax, VA 22030-4444, USA}

\and

\author{Michael T. Wolff\altaffilmark{3}}
\affil{E. O. Hulburt Center for Space Research \break
Naval Research Laboratory, \break
Washington, DC 20375, USA}

\vfil

\altaffiltext{1}{pbecker@gmu.edu}
\altaffiltext{2}{also Department of Physics and Astronomy,
George Mason University, Fairfax, VA 22030-4444, USA}
\altaffiltext{3}{michael.wolff@nrl.navy.mil}

\begin{abstract}
Accretion-powered X-ray pulsars are among the most luminous X-ray
sources in the Galaxy. However, despite decades of theoretical and
observational work since their discovery, no satisfactory model for the
formation of the observed X-ray spectra has emerged. In particular, the
previously available theories are unable to reproduce the power-law
variation observed at high energies in many sources. In this paper, we
present the first self-consistent calculation of the spectrum emerging
from a pulsar accretion column that includes an explicit treatment of
the energization occurring in the shock. Using a rigorous eigenfunction
expansion method based on the exact dynamical solution for the velocity
profile in the column, we obtain a closed-form expression for the
Green's function describing the upscattering of radiation injected into
the column from a monochromatic source located at the top of the thermal
mound, near the base of the flow. The Green's function is convolved with
a Planck distribution to calculate the radiation spectrum resulting from
the reprocessing of blackbody photons emitted by the thermal mound. We
demonstrate that the energization of the photons in the shock naturally
produces an X-ray spectrum with a power-law shape at high energies and a
blackbody shape at low energies, in agreement with many observations of
accreting X-ray pulsars.

\end{abstract}



\section{INTRODUCTION}

Since the discovery of the first known pulsating X-ray sources Her X-1
and Cen X-3 more than three decades ago (Giacconi et al. 1971; Tananbaum
et al. 1972), over 50 new sources have been detected in the galaxy and
the Magellanic Clouds, with luminosities in the range $L_{\rm X} \sim
10^{34-38}{\ \rm ergs \ s^{-1}}$ and pulsation periods $0.1 {\ \rm s}
\lapprox P \lapprox 10^3 {\ \rm s}$. X-ray pulsars include a variety of
objects powered by rotation or accretion, as well as several anomalous
X-ray pulsars whose energy source is currently unclear. The emission
from X-ray pulsars in binary systems is fueled by the accretion of
material from the ``normal'' companion onto the neutron star, with the
flow channeled onto one or both of the magnetic poles by the strong
field. During the accretion process, gravitational potential energy is
converted into kinetic energy, which escapes from the column in the form
of X-rays as the gas decelerates through a radiative shock before
settling onto the stellar surface. In the accretion-powered sources,
which are the focus of this paper, the X-ray spectra are often well
fitted using a combination of a power-law spectrum plus a blackbody
component with a temperature in the range $T \sim 10^{6-7}\,$K (e.g.,
Coburn et al. 2002; di Salvo et al. 1998; White et al. 1983). Most
spectra also display quasi-exponential cutoffs at $E \sim 20-30\,$keV,
and there are indications of cyclotron features and iron emission lines
in a number of sources. The observations suggest typical magnetic
field strengths of $\sim 10^{12-13}\,$G.

Although accretion-powered X-ray pulsars are among the most luminous
sources in the galaxy, previous attempts to calculate their spectra
based on static or dynamic theoretical models have generally yielded
results that do not agree very well with the observed profiles (e.g.,
M\'esz\'aros \& Nagel 1985a,b; Nagel 1981; Yahel 1980; Klein et al.
1996). Hence there is still no clear understanding of the basic spectral
formation mechanism in X-ray pulsars (see the discussion in Coburn et
al. 2002). Given the lack of a viable theoretical model, X-ray pulsar
spectral data have traditionally been fit using multicomponent forms
that include absorbed power-laws, cyclotron features, iron emission
lines, blackbody components, and high-energy exponential cutoffs. The
resulting parameters are sometimes difficult to relate to the physical
properties of the source. Motivated by the lack of a comprehensive
theoretical model for X-ray pulsar spectral formation, we reconsider
here the physical picture originally proposed by Davidson (1973), in
which the accreting gas passes through a radiative, radiation-dominated
shock before settling onto the surface of the star. We illustrate the
accretion/emission geometry schematically in Figure~1. Most of the
photons emitted from the accretion column are produced in the dense
``thermal mound'' located inside the column, just above the stellar
surface. The blackbody photons created in the mound are upscattered in
the shock, and eventually diffuse through the walls of the column. The
escaping photons carry away the kinetic energy of the gas, thereby
allowing the plasma to settle onto the surface of the star.

\subsection{Bulk Comptonization}

The strong compression that occurs as the plasma crosses the radiative
shock renders it an ideal site for first-order Fermi energization (i.e.,
``bulk'' or ``dynamical'' Comptonization) of the photons produced by the
thermal mound. In the bulk Comptonization process, particles experience
a mean energy gain if the scattering centers they collide with are
involved in a converging flow (e.g., Laurent \& Titarchuk 1999; Turolla,
Zane, \& Titarchuk 2002). By contrast, in the thermal Comptonizaton
process, particles gain energy due to the {\it stochastic} motions of
the scattering centers via the second-order Fermi mechanism (e.g.,
Sunyaev \& Titarchuk 1980; Becker 2003). In the X-ray pulsar
application, the scattering centers are infalling electrons, and the
energized ``particles'' are photons. Since the inflow speed of the
electrons in an X-ray pulsar accretion column is much larger than their
thermal velocity, bulk Comptonization dominates over the stochastic
process except at the highest photon energies (Titarchuk, Mastichiadis,
\& Kylafis 1996), as discussed in \S~7. The failure of the current
models to generate the power-law spectral shape characteristic of X-ray
pulsars probably stems from the neglect of the critical contribution of
the shock in upscattering the soft radiation produced in the thermal
mound (e.g., Burnard, Arons, \& Klein 1991). We demonstrate in this
paper that bulk Comptonization of the radiation in the accretion shock
can produce spectra very similar to the power-law continuum seen in many
X-ray pulsars. The main results presented here were summarized by Becker
\& Wolff (2005).

Dynamical Comptonization has been considered by a number of previous
authors in a variety of astrophysical situations, including spherical
inflows (Payne \& Blandford 1981; Colpi 1988; Schneider \& Bogdan 1989;
Titarchuk, Mastichiadis, \& Kylafis 1997), accretion disks around black
holes (Laurent \& Titarchuk 2001; Titarchuk, Cui, \& Wood 2002;
Titarchuk \& Shrader 2002), spherical outflows (Becker \& Begelman 1986;
Titarchuk, Kazanas, \& Becker 2003), and the transport of cosmic rays in
the solar wind (Parker 1965; Stawicki, Fichtner, \& Schlickeiser 2000).
The process was also studied in the context of neutron star accretion by
Titarchuk, Mastichiadis, \& Kylafis (1996) and Mastichiadis \& Kylafis
(1992), although their results are not applicable to X-ray pulsar
accretion columns due to the assumption of spherical symmetry and the
imposition of a power-law velocity profile. Lyubarskii \& Sunyaev (1982)
analyzed dynamical Comptonization in the context of plane-parallel
pulsar shocks, but the velocity profile they utilized is not consistent
with the dynamics of X-ray pulsar accretion, and furthermore they did
not account for the escape of radiation through the walls of the column.
The relevance of the bulk Comptonization process for spectral formation
in X-ray pulsars was also recognized by Burnard, Arons, \& Klein (1991),
who pointed out that thermal spectra alone are much too soft to explain
the observed emission, and suggested that upscattering in the shock
could provide the required hardening.

The fundamental character of the Green's function describing both
thermal and bulk Comptonization was studied by Titarchuk \& Zannias
(1998) for the case of accretion onto a black hole. These authors
established that the Green's function can be approximated using a broken
power-law form with a central peak between the high- and low-energy
portions of the spectrum, for either type of Comptonization.
Furthermore, they concluded that bulk Comptonization dominates over the
thermal process if the electron temperature $T \lapprox 10^7\,$K. The
theory developed in the present paper represents an extension of the
same idea to neutron star accretion, resulting in the exact Green's
function for the X-ray pulsar spectral formation process. In agreement
with Titarchuk \& Zannias (1998), we find that the bulk Comptonization
process is dominant for the temperature range relevant for X-ray
pulsars. However, the presence of the event horizon in the black-hole
application treated by these authors makes their problem fundamentally
different from the neutron star accretion problem studied here, because
the neutron star obviously possesses a solid surface. Our work therefore
represents the first exact, quantitative analysis of the role of bulk
Comptonization in the X-ray pulsar spectral formation process.

\subsection{Radiation-Dominated Flow}

Radiation pressure governs the dynamical structure of the accretion
flows in bright pulsars when the X-ray luminosity satisfies (Becker
1998; Basko \& Sunyaev 1976)
\begin{equation}
\xlum \sim L_{\rm crit} \equiv
{2.72 \times 10^{37} \sig \over
\sqrt{\sigperp\sigpar}}
\left(M_* \over \msun \right)
\left(\colrad \over R_*\right)
\ {\rm \ ergs \ s}^{-1}
\ ,
\label{eq1.1}
\end{equation}
where $\colrad$ is the polar cap radius, $\starmass$ and $\starad$
denote the stellar mass and radius, respectively, $\sig$ is the Thomson
cross section, and $\sigpar$ and $\sigperp$ represent the electron
scattering cross sections for photons propagating parallel or
perpendicular to the magnetic field, respectively. When the luminosity
of the system is comparable to $L_{\rm crit}$, the radiation flux in the
column is super-Eddington and therefore the radiation pressure greatly
exceeds the gas pressure (Becker 1998). In this situation the gas passes
through a radiation-dominated shock on its way to the stellar surface,
and the kinetic energy of the gas is carried away by the high-energy
radiation that escapes from the column. The strong gradient of the
radiation pressure decelerates the material to rest at the surface of
the star. The observation of many X-ray pulsars with $L_{\rm X} \sim
10^{36-38} \, {\rm ergs \, s}^{-1}$ implies the presence of
radiation-dominated shocks close to the stellar surfaces in these
systems (White et al. 1983; White et al. 1995). Note that
radiation-dominated shocks are {\it continuous} velocity transitions,
with an overall thickness of a few Thomson scattering lengths, unlike
traditional (discontinuous) gas-dominated shocks (Blandford \& Payne
1981b).

In luminous X-ray pulsars, the pressure of the radiation determines the
dynamics of the flow, while the dynamics in turn determines the shape of
the radiation spectrum. Hence the flow dynamics and the radiative
transport are intimately connected, which makes this a complex and
nonlinear ``photohydrodynamical'' problem. One implication is that the
photons scattering through the column cannot be regarded as ``test
particles'' since it is their pressure that dictates the structure of
the flow. We must therefore solve for the radiation spectrum and the
velocity profile in a self-consistent manner. Radiation pressure may
also play an important role in the dynamics of moderate-luminosity pulsars
due to the strong dependence of the electron scattering (cyclotron)
cross section on the magnetic field strength (Langer \& Rappaport 1982).

Becker (1998) and Basko \& Sunyaev (1976) have considered the dynamics
of radiation-dominated pulsar accretion flows, including the effect of
the shock and the escape of radiation energy from the column. They find
that the gas is decelerated to zero velocity in a manner consistent with
accretion onto a solid body, but they do not consider the shape of the
escaping radiation spectrum. Conversely, the dynamics and the radiation
spectrum have been worked out self-consistently by Blandford \& Payne
(1981b) for the case of a ``standard'' (Rankine-Hugoniot)
radiation-dominated shock, which does not include radiation escape. In
the Blandford \& Payne approach, the radiation spectrum everywhere in
the flow is determined by solving the transport equation using the
incident radiation field as a boundary condition. The solutions have a
power-law character at high energies, with a spectral index that depends
on the Mach number of the upstream flow. Since the Rankine-Hugoniot
shocks studied by Blandford \& Payne have a conserved energy flux, the
downstream velocity is never less than one-seventh its upstream value.
Consequently the associated spectral solutions are not applicable to the
X-ray pulsar case, since pulsar shocks must be {\it radiative} in nature
so that the kinetic energy of the infalling material can be removed and
the flow brought to rest at the stellar surface.

The analytical approach employed here parallels the work of Blandford \&
Payne (1981b), except that the velocity profile they utilized is
replaced with the appropriate solution for an X-ray pulsar accretion
column, and the loss of radiation energy from the column is incorporated
into the transport equation using an escape-probability formalism. In
the spirit of the Blandford \& Payne study of spectral formation in
radiation dominated Rankine-Hugoniot shocks, we are interested here in
exploring the direct effect of the radiative accretion shock on the
spectrum emerging from an X-ray pulsar. Hence it is not our goal in this
paper to develop complete models that include additional processes such
as free-free emission and absorption, cyclotron features, and iron
emission lines. However, even without including any of these processes,
we are able to demonstrate qualitative agreement with X-ray pulsar
spectra. This suggests that energization in the accretion shock is one
of the most important aspects of spectral formation in X-ray pulsars.

\begin{figure}[t]
\begin{center}
\epsfig{file=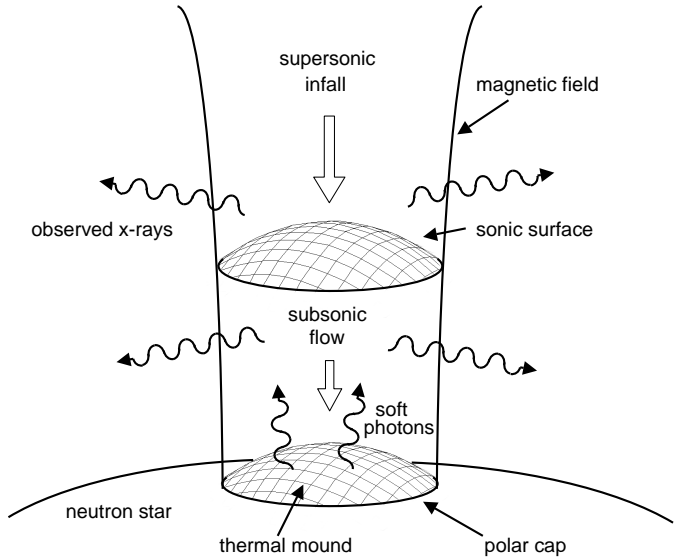,height=15.0cm,angle=270}
\end{center}
\vskip-1.0truein
\caption{Schematic depiction of gas accreting onto one of the
magnetic polar caps of a neutron star.}
\end{figure}

The remainder of the paper is organized as follows. In \S~2 we briefly
review the conservation equations and discuss the resulting dynamical
solution for the velocity profile of the accreting gas. The transport
equation governing the scattering of the radiation inside the accretion
column is analyzed in \S~3, and various constraints are derived in order
to ensure that the radiative transfer is modeled in a manner consistent
with the flow dynamics. In \S~4 we obtain the exact analytical solution
for the Green's function and convolve it with the Planck distribution
emitted at the thermal mound to compute the radiation spectrum emerging
from the accretion column. In \S~5 we present results obtained for the
count-rate spectra using parameters corresponding to specific pulsars,
and we compare the theoretical spectra with the observational data. The
implications of our results for the production of X-ray spectra in
pulsars are discussed in \S~7. Additional mathematical details are
provided in a series of Appendices.

\section{DYNAMICS OF RADIATION-DOMINATED FLOW}

The model analyzed here is based on the dynamical picture suggested by
Davidson (1973), which is illustrated schematically in Figure~1. The
accretion scenario corresponds physically to the flow of a mixture of
gas and radiation inside a magnetic ``pipe'' that is sealed with
respect to the gas, but is transparent with respect to the radiation.
The accretion column incorporates a radiation-dominated, radiative shock
located above the stellar surface. Although the shock is extended and
the velocity profile is continuous, the flow possesses a well-defined
sonic surface, as indicated in Figure~1. The model also includes a dense
thermal mound located at the base of the flow, where local thermodynamic
equilibrium prevails. Due to its blackbody nature, the thermal mound
acts as both a photon source (emitting a Planck spectrum) and a photon
sink (absorbing all incident radiation). Hence the surface of the mound
represents the ``photosphere'' for photon creation and absorption, and
the opacity is dominated by electron scattering above this point.

In our approach to the problem, we assume that the upstream flow is
composed of pure, fully ionized hydrogen gas moving at a highly
supersonic speed, which is the standard scenario for accretion-powered
X-ray pulsars (Basko \& Sunyaev 1975, 1976). Our transport model employs
a cylindrical, plane-parallel geometry, and therefore the velocity,
density, and pressure are functions of the distance above the stellar
surface, but they are all constant across the column at a given height.
In this situation, the vertical variation of the flow velocity is given
by the exact analytical solution obtained by Becker (1998) and Basko \&
Sunyaev (1976), which describes the settling of the transonic flow onto
the surface of the star. A detailed consideration of the angular and
energy dependences of the electron scattering cross section is beyond
the scope of the present paper. We shall therefore follow Wang \& Frank
(1981) and Becker (1998) by treating the directional dependence of the
electron scattering in an approximate way in terms of the
energy-averaged cross sections $\sigpar$ and $\sigperp$, describing
respectively the scattering of photons propagating either parallel or
perpendicular to the magnetic field. The gas in the accretion column is
radiation-dominated, and therefore it is the pressure of the photons
that decelerates the infalling plasma to rest. We will use this fact in
\S~6.1 to evaluate the self-consistency of the coupled radiative
transport/dynamical model based on the analytical solution for the
radiation pressure profile obtained below.

\subsection{Conservation Equations}

The one-dimensional, time-dependent Euler equations governing the flow
of a radiation-dominated gas with mass density $\rho$, flow velocity
$v$, and photon energy density $U$ inside a cylindrical accretion column
are
\begin{equation}
{\partial\rho\over\partial t} = -{\partial J\over\partial x}
\ ,
\label{eq2.1}
\end{equation}
\begin{equation}
{\partial\over\partial t}\left(\rho v\right)
= -{\partial I\over\partial x}
\ ,
\label{eq2.2}
\end{equation}
\begin{equation}
{\partial\over\partial t}\left({1\over 2}\,\rho v^2+U\right)
= -{\partial E\over\partial x} + \dot U_{\rm esc}
+ \dot U_{\rm abs} + \dot U_{\rm emit}
\ ,
\label{eq2.3}
\end{equation}
where the mass, momentum, and energy fluxes are given
respectively by
\begin{equation}
J = \rho \, v
\ ,
\label{eq2.4}
\end{equation}
\begin{equation}
I = P+\rho \, v^2
\ ,
\label{eq2.5}
\end{equation}
\begin{equation}
E = {1\over 2} \, \rho v^3 + P \, v + U \, v
- c \, {\partial P\over\partial\tau_\|}
\ ,
\label{eq2.6}
\end{equation}
with $P=U/3$ denoting the radiation pressure. The spatial coordinate $x$
increases along the column axis, in the direction of the flow, and
$\taupar$ denotes the associated electron-scattering optical depth,
which is related to $x$ via
\begin{equation}
d\tau_\| = n_e \, \sigpar \, dx
\ ,
\label{eq2.7}
\end{equation}
where $n_e = \rho / m_p$ is the electron number density, $m_p$ is the proton
mass, and $\sigpar$ is the electron scattering cross section parallel to
the magnetic field (i.e., the column axis). The two coordinates $x$ and
$\taupar$ are calibrated so that they each vanish at the sonic point, as
explained below.

The quantities $\dot U_{\rm esc}$, $\dot U_{\rm abs}$, and $\dot U_{\rm
emit}$ appearing in equation~(\ref{eq2.3}) represent the rates per unit
volume at which radiation energy escapes through the column walls, or is
absorbed or emitted by the gas, respectively. When the system is
radiation-dominated as assumed here, the gas itself has little thermal
energy to contribute to the radiation field, although it does possess a
large reservoir of {\it bulk} kinetic energy which is transferred to the
photons through the first-order Fermi energization process in the shock.
In this situation, the emission and absorption processes must nearly
balance throughout the flow, and therefore we can write
\begin{equation}
\dot U_{\rm abs} + \dot U_{\rm emit} = 0
\ ,
\label{eq2.8}
\end{equation}
in which case equation~(\ref{eq2.3}) reduces in a steady state to
\begin{equation}
{\partial E\over\partial x} = \dot U_{\rm esc}
\ .
\label{eq2.9}
\end{equation}
Although emission and absorption do not effect the flow dynamics in this
situation, these processes profoundly influence the {\it shape} of the
radiation spectrum, as discussed in \S\S~3 and~4.

Within the context of our one-dimensional picture, the rate at which
radiation energy diffuses through the walls of the column can be
approximated using an escape-probability formalism. If the column cross
section is optically thick to electron scattering, as expected, then the
rate at which radiation energy escapes through the walls is given by
\begin{equation}
\dot U_{\rm esc} = - \, {U \over t_{\rm esc}}
\ ,
\label{eq2.10}
\end{equation}
where the mean escape time, $t_{\rm esc}$, is given by
\begin{equation}
t_{\rm esc} = {r_0 \over w_\perp} \ , \ \ \ \ \ 
w_\perp = {c \over \tauperp} \ , \ \ \ \ \ 
\tauperp = n_e \, \sigperp \, \colrad
\ ,
\label{eq2.11}
\end{equation}
with $w_\perp$ denoting the diffusion velocity perpendicular to the
$x$-axis, and $\tauperp$ representing the perpendicular optical
thickness of the cylindrical accretion column with radius $r_0$. Note
that $\tauperp$ and $t_{\rm esc}$ are functions of $x$ through their
dependence on $n_e$. Becker (1998) confirmed that the diffusion
approximation employed in equation~(\ref{eq2.10}) is valid because
$\tauperp > 1$ for typical X-ray pulsar parameters.
Equations~(\ref{eq2.4}) and (\ref{eq2.11}) can be combined to express
the mean escape time as
\begin{equation}
t_{\rm esc} = {J \, r_0^2 \, \sigperp \over m_p \, c \, v}
\ .
\label{eq2.12}
\end{equation}
Since the escape timescale is inversely proportional to the flow
velocity, the column becomes completely opaque at the surface of the
neutron star due to the divergence of the electron number density there.
The relationship between the escape-probability approximation and the
physical distribution of radiation inside the accretion column is further
discussed in \S~6.3.

\subsection{Dynamical Solution}

In the steady-state situation of interest here, the mass flux $J$ and
the momentum flux $I$ are both conserved, but the energy flux $E$ decreases
as the gas accretes onto the surface of the neutron star due to the
emission of radiation through the column walls. The momentum, energy,
and mass conservation equations can be combined in this case to show
that the flow velocity $v$ satisfies the second-order nonlinear
differential equation (Becker 1998)
\begin{equation}
{d \over d\tau}\left(-{7 \over 2} \, \mu^2 + 7 \, \mu
+ {d \mu \over d\tau}\right) = - \mu^2 \, (7  - 4 \, \mu)
\ ,
\label{eq2.13}
\end{equation}
where
\begin{equation}
\tau \equiv \tau_\| \, {v_c \over c} \ , \ \ \ \ \ 
\mu \equiv {v \over v_c}
\ .
\label{eq2.14}
\end{equation}
Here, $v_c$ represents the flow velocity at the sonic point, where
we set $x=\tau=0$ and ${\cal M}=1$, with the radiation Mach number
${\cal M}$ defined by
\begin{equation}
{\cal M} \equiv {v \over c_s} \ , \ \ \ \ \ 
c_s^2 \equiv {4 \over 3} \, {P \over \rho}
\ .
\label{eq2.15}
\end{equation}
The quantity $c_s$ denotes the speed of sound in the radiation-dominated
gas. In the far upstream region, we assume that ${\cal M} \to \infty$
which is an excellent approximation in pulsar accretion flows (Basko
\& Sunyaev 1975, 1976).

Becker (1998) showed that in order for the flow to come to rest at the
stellar surface, the parameters $r_0$, $J$, $\sigperp$, and $\sigpar$
must satisfy the relation
\begin{equation}
{m_p^2 \, c^2 \over r_0^2 \, J^2 \, \sigperp \, \sigpar}
= {4 \over 3}
\ .
\label{eq2.16}
\end{equation}
The corresponding exact solution for the velocity $v$ as a function of
$\tau$ in the steady-state situation of interest here is given by
\begin{equation}
v(\tau) = {7 \, v_c \over 8} \left\{1 - \tanh\left[{7 \over 2} \,
(\tau-\tau_*)\right]\right\}
\ ,
\label{eq2.17}
\end{equation}
where
\begin{equation}
\tau_* \equiv {2 \over 7} \, \tanh^{-1}\left(1 \over 7\right)
\approx 0.041
\ .
\label{eq2.18}
\end{equation}
Note that $\mu=1$ at the sonic point ($\tau=0$) as required. The
velocity can also be written as an explicit function of the position $x$
using (Basko \& Sunyaev 1976)
\begin{equation}
v(x) = {7 \, v_c \over 4} \, \left[1 - \left(7 \over 3\right)
^{-1+x/x_{\rm st}}\right]
\ ,
\label{eq2.19}
\end{equation}
where $x_{\rm st}$ is the distance between the sonic point and the
stellar surface, which can be evaluated using equation~(4.16) from
Becker (1998) to obtain
\begin{equation}
x_{\rm st} = {r_0 \over 2 \sqrt{3}} \left(\sigperp \over \sigpar
\right)^{1/2} \ln\left(7 \over 3\right)
\ .
\label{eq2.20}
\end{equation}
Note that $x$ increases in the direction of the flow, which comes to rest
at the surface of the star as expected. The height above the stellar surface
$h$ is related to the coordinate $x$ via
\begin{equation}
h(x) = x_{\rm st} - x
\ .
\label{eq2.21}
\end{equation}
The incident (upstream) velocity of the infalling material is expected
to be close to the free-fall velocity onto the stellar surface, $\vff$,
given by
\begin{equation}
\vff = \left(2 \, G M_* \over R_*\right)^{1/2}
\ .
\label{eq2.22}
\end{equation}
Since $v \to (7/4) \, v_c$ in the upstream region according to
equation~(\ref{eq2.19}), it follows that $v_c = (4/7) \, \vff$,
and therefore the flow velocity at the sonic point is given by
\begin{equation}
v_c = {4 \over 7} \left(2 \, G M_* \over R_*\right)^{1/2}
\ .
\label{eq2.23}
\end{equation}

With the velocity solution given by equation~(\ref{eq2.19}), the
corresponding pressure profile can be obtained by combining
equations~(\ref{eq2.4}) and (\ref{eq2.5}), which yields
\begin{equation}
P(x) = I - J \, v(x)
\ .
\label{eq2.24}
\end{equation}
The pressure increases to a maximum value at the surface
of the star, where $v \to 0$. Based on equations~(\ref{eq2.4}) and
(\ref{eq2.15}), we can reexpress the Mach number as
\begin{equation}
{\cal M} = \left({3 \, J \, v \over 4 \, P}\right)^{1/2}
\ .
\label{eq2.24b}
\end{equation}
Since the Mach number approaches infinity in the far upstream region and
$v \to (7/4) \, v_c$, it follows that $P$ must vanish asymptotically
and consequently the upstream momentum flux is dominated by the ram
pressure of the freely falling matter. We therefore obtain
\begin{equation}
I = J \, \vff
\ .
\label{eq2.25}
\end{equation}
By combining equations~(\ref{eq2.23}), (\ref{eq2.24}) and
(\ref{eq2.25}), we find that the pressure profile is related
to the velocity variation by
\begin{equation}
P(x) = J \, v_c \left[{7 \over 4} - {v(x) \over v_c}\right]
\ .
\label{eq2.26}
\end{equation}
Using equation~(\ref{eq2.19}) to eliminate the velocity $v(x)$
yields the closed-form result
\begin{equation}
P(x) = {7 \over 4} \, J \, v_c \left(7 \over 3\right)
^{-1+x/x_{\rm st}}
\ .
\label{eq2.27}
\end{equation}
The pressure vanishes in the upstream limit ($x \to -\infty$) as expected,
and at the surface of the star ($x = x_{\rm st}$), the pressure achieves
the stagnation value
\begin{equation}
P_{\rm st} \equiv P(x_{\rm st}) = {7 \over 4} \, J \, v_c
\ .
\label{eq2.28}
\end{equation}

\section{RADIATIVE TRANSFER IN THE ACCRETION COLUMN}

If the gas is radiation-dominated and fully ionized, then the photons
interact with the matter primarily via electron scattering, which
controls both the spatial transport and the energization of the
radiation. In this situation the opacity is dominated by electron
scattering, and therefore absorption is negligible, except at the surface
of the thermal mound located inside the cylindrical accretion column at
$x=x_0$. We are interested in obtaining the photon distribution,
$f(x_0,x,\epsilon)$, measured at position $x$ and energy $\epsilon$
resulting from the reprocessing of blackbody radiation emitted from the
mound. The normalization of $f$ is defined so that $\epsilon^2
f(x_0,x,\epsilon) \, d\epsilon$ gives the number density of photons in
the energy range between $\epsilon$ and $\epsilon+d\epsilon$, and
therefore $f = 8 \pi \, \bar n / (c^3 h^3)$, where $\bar n$ is the
occupation number. The self-consistency of the solution obtained
for the radiation distribution $f$ will be confirmed by verifying
that the associated pressure distribution agrees with the dynamical
result given by equation~(\ref{eq2.27}).

\subsection{Transport Equation}

In the cylindrical geometry employed here, the photon distribution $f$
satisfies the transport equation (e.g., Blandford \& Payne 1981a; Becker
1992; Parker 1965; Becker \& Wolff 2005)
\begin{equation}
{\partial f \over \partial t} = - v \, {\partial f \over \partial x}
+ {dv\over d x}\,{\epsilon\over 3} \,
{\partial f\over\partial\epsilon} + {\partial\over\partial x}
\left({c\over 3 n_e \sigma_\|}\,{\partial f\over\partial x}\right)
+ {S(\epsilon) \over \pi r_0^2} \, \delta(x-x_0)
- {f \over t_{\rm esc}}
- \beta \, v_0 \, \delta(x-x_0) \, f
\ ,
\label{eq3.1}
\end{equation}
where $\epsilon$ is the photon energy, $x$ is the spatial coordinate
along the column axis, and $v_0 \equiv v(x_0)$ is the flow velocity at
the top of the thermal mound. The terms on the right-hand side of
equation~(\ref{eq3.1}) represent advection, first-order Fermi
energization (``bulk Comptonization'') in the converging flow, spatial
diffusion parallel to the column axis, the blackbody source, escape of
radiation from the column, and the absorption of radiation at the
thermal mound, respectively. We are interested here in the steady-state
version of equation~(\ref{eq3.1}) with $\partial f/\partial t=0$. The
photon number and energy densities associated with the distribution $f$
are given respectively by
\begin{equation}
n(x) = \int_0^\infty \epsilon^2 \,
f(x_0,x,\epsilon) \, d\epsilon
\ , \ \ \ \ 
U(x) = \int_0^\infty \epsilon^3 \,
f(x_0,x,\epsilon) \, d\epsilon
\ .
\label{eq3.2}
\end{equation}
In the present application, we are primarily interested in exploring the
effect of the dynamics itself (i.e., the radiative shock) on the
observed X-ray emission. The transport equation~(\ref{eq3.1}) therefore
does not include additional processes such as thermal Comptonization or
cyclotron emission and absorption that are likely to be important in
X-ray pulsars. In general, the neglect of thermal Comptonization is
reasonable because at the temperatures typical of X-ray pulsars ($T \sim
10^{6-7}\,$K), the kinetic energy associated with the bulk flow far
surpasses the thermal energy. However, the failure to include the
electron recoil associated with thermal Comptonization renders
the model unable to reproduce the quasi-exponential cutoffs often
observed in X-ray pulsar spectra at high energies. Furthermore, in some
sources thermal Comptonization appears to be necessary in order to
flatten the spectrum in the energy range $\epsilon \sim 5-20\,$keV. We
provide further discussion of this issue in \S~7.

When integrated over the photon energy spectrum, the first-order Fermi
process considered here corresponds to the $PdV$ work done on the
radiation by the compression of the background plasma as it accretes
onto the stellar surface. The mean escape time, $t_{\rm esc}$, is given
by equation~(\ref{eq2.12}), and the source term $S(\epsilon)$ is defined
so that $\epsilon^2 S(\epsilon) \, d\epsilon$ represents the number of
photons injected into the accretion column per second in the energy
range between $\epsilon$ and $\epsilon + d\epsilon$ from the blackbody
surface at $x=x_0$. The absorption term in equation~(\ref{eq3.1})
containing the dimensionless constant $\beta$ must be included due to
the blackbody nature of the thermal mound, which acts as both a source
and a sink of radiation. In the radiation-dominated situation considered
here, essentially all of the pressure is provided by the photons, and
therefore the rate of absorption of radiation energy at the top of the
mound must equal the energy emission rate when integrated with respect
to the photon frequency (see eq.~[\ref{eq2.8}]). We will use this energy
balance argument to compute a self-consistent value for $\beta$ below.

Next we need to specify the appropriate form for the source term
$S(\epsilon)$ in order to treat the injection of the blackbody radiation
from the surface of the thermal mound. The flux per unit energy measured
at the surface of an isotropically radiating object is equal to $\pi$
times the intensity (Rybicki \& Lightman 1979). It therefore follows
that the amount of energy emitted per second from the upper surface of
the thermal mound (with area $\pi r_0^2$) in the energy range between
$\epsilon$ and $\epsilon + d\epsilon$ is given by
\begin{equation}
\epsilon^3 S(\epsilon) \, d\epsilon
= \pi \, r_0^2 \cdot \pi \, B_\epsilon(\epsilon) \, d\epsilon \ ,
\label{eq3.3}
\end{equation}
where
\begin{equation}
B_\epsilon(\epsilon) = {2 \, \epsilon^3 \over h^3 \, c^2} \, {1
\over e^{\epsilon/kT_0} - 1}
\label{eq3.4}
\end{equation}
denotes the blackbody intensity and $T_0 \equiv T(x_0)$ is the gas
temperature at the surface of the mound. Note that the units for
$B_\epsilon$ are ${\rm ergs \ s^{-1} \, ster^{-1} \,
cm^{-2} \, erg^{-1}}$. Hence we obtain for the source term
\begin{equation}
S(\epsilon) = {2 \, \pi^2 \, r_0^2 \over h^3 \, c^2}
\, {1 \over e^{\epsilon/kT_0} - 1}
\ ,
\label{eq3.5}
\end{equation}
and consequently the total energy injection rate is given by
\begin{equation}
\int_0^\infty \epsilon^3 \, S(\epsilon) \, d\epsilon
= \pi \, r_0^2 \ \sigma \, T_0^4 \ \ \ \propto \ {\rm ergs \ s^{-1}}
\ ,
\label{eq3.6}
\end{equation}
where $\sigma$ is the Stephan-Boltzmann constant.

\subsection{Energy Balance at the Thermal Mound}

In a radiation-dominated X-ray pulsar accretion column, the emission of
fresh blackbody radiation energy at the surface of the thermal mound is
almost perfectly balanced by the absorption of energy, as expressed by
equation~(\ref{eq2.8}). Most of the energy appearing in the emergent
X-rays is therefore not supplied from the internal energy of the gas,
but rather from its bulk kinetic energy, which is transferred directly
to the photons via collisions with infalling electrons. We can gain some
insight into the energy balance at the thermal mound by rewriting the
steady-state version of the transport equation~(\ref{eq3.1}) in the
flux-conservation form (Skilling 1975; Gleeson \& Axford 1967)
\begin{equation}
{\partial H_\epsilon \over \partial x} = - \, {1 \over 3 \epsilon^2}
{\partial \over \partial \epsilon} \left(\epsilon^3 v \, {\partial f \over
\partial x} \right)
+ {S(\epsilon) \over \pi r_0^2} \, \delta(x-x_0)
- {f \over t_{\rm esc}}
- \beta \, v_0 \, \delta(x-x_0) \, f \ ,
\label{eq3.7}
\end{equation}
where
\begin{equation}
H_\epsilon \equiv - \, {c \over 3 \, n_e \, \sigpar} \,
{\partial f\over \partial x} - {v \, \epsilon \over 3} \,
{\partial f \over \partial \epsilon}
\label{eq3.8}
\end{equation}
represents the ``specific flux'' (Becker 1992). By operating on
equation~(\ref{eq3.7}) with $\int_0^\infty \epsilon^3 \, d\epsilon$
and utilizing equations~(\ref{eq3.2}) and (\ref{eq3.6}), we arrive
at the photon energy equation
\begin{equation}
{dQ \over dx}
= {v \over 3} \, {dU \over dx}
- {U \over t_{\rm esc}} - \beta \, v_0 \, \delta(x-x_0) \, U
+ \sigma \, T_0^4 \, \delta(x-x_0)
\ ,
\label{eq3.9}
\end{equation}
where the integrated photon energy flux, $Q$, is given by
\begin{equation}
Q \equiv \int_0^\infty \epsilon^3 \, H_\epsilon \, d\epsilon
= {4 \over 3} \, v \, U - {c \over 3
\, n_e \, \sigpar} \, {dU \over dx}
\ \ \propto \ \ {\rm ergs \ cm^{-2} \ s^{-1}}
\ .
\label{eq3.10}
\end{equation}
Equation~(\ref{eq3.9}) states that the divergence of the radiation
energy flux $Q$ is equal to the net rate of change of the photon energy
density $U$ due to the combined influence of compression, escape, absorption,
and photon injection. Note that $Q$ is related to the {\it total} energy
flux $E$ via (see eq.~[\ref{eq2.6}])
\begin{equation}
E = Q + {1\over 2}\,\rho v^3
\ .
\label{eq3.11}
\end{equation}
Since the absorption and emission of radiation energy must balance at
the surface of the mound in the radiation-dominated situation of
interest here, we can integrate equation~(\ref{eq3.9}) with respect to
$x$ across the mound location to obtain
\begin{equation}
\lim_{\varepsilon \to 0} \ Q(x_0+\varepsilon)
- Q(x_0-\varepsilon) = - \beta \, v_0 \, U_0 + \sigma \, T_0^4 = 0
\ ,
\label{eq3.12}
\end{equation}
where $U_0 \equiv U(x_0)$ denotes the radiation energy density at the
mound surface. The fact that the photon energy flux $Q$ is continuous at
$x=x_0$ therefore requires that the dimensionless absorption constant
$\beta$ in the transport equation~(\ref{eq3.1}) be given by
\begin{equation}
\beta = {\sigma \, T_0^4 \over v_0 \, U_0}
\ .
\label{eq3.13}
\end{equation}
By utilizing this expression for $\beta$, we ensure that the results
obtained for the radiation spectrum $f(x_0,x,\epsilon)$ are consistent
with the dynamical structure of the accretion column. This is confirmed
after the fact by calculating the radiation pressure $P = U/3$ using
equation~(\ref{eq3.2}) and comparing the result with the dynamical
solution given by equation~(\ref{eq2.27}).

\section{EXACT SOLUTION FOR THE RADIATION SPECTRUM}

In our approach to solving equation~(\ref{eq3.1}) for the spectrum
$f(x_0,x,\epsilon)$ inside the accretion column, we shall first obtain
the Green's function, $\green(x_0,x,\epsilon_0,\epsilon)$, which is the
radiation distribution at location $x$ and energy $\epsilon$ resulting
from the injection of $\dot N_0$ photons per second with energy
$\epsilon_0$ from a monochromatic source at location $x_0$. The
determination of the Green's function is a useful intermediate step in
the process because it provides us with fundamental physical insight
into the spectral redistribution process, and it also allows us to
calculate the particular solution for the spectrum $f$ associated with
an arbitrary continuum source $S(\epsilon)$ using the integral
convolution (Becker 2003)
\begin{equation}
f(x_0,x,\epsilon) = \int_0^\epsilon {\green(x_0,x,\epsilon_0,\epsilon)
\over \dot N_0} \ \epsilon_0^2 \, S(\epsilon_0) \, d\epsilon_0
\ .
\label{eq4.1}
\end{equation}
Note that the upper bound of integration is set equal to $\epsilon$
because in our model all of the injected photons gain energy via the
first-order Fermi process, and none lose energy. The technical procedure
used to solve for the Green's function involves the derivation of
eigenfunctions and associated eigenvalues based on the set of spatial
boundary conditions for the problem (see, e.g., Blandford \& Payne
1981b; Payne \& Blandford 1981; Schneider \& Kirk 1987; Colpi 1988).

The steady-state transport equation governing the Green's function is
(cf. eq.~[\ref{eq3.1}])
\begin{equation}
v{\partial\green\over\partial x}={dv\over d x}\,{\epsilon\over 3}\,
{\partial\green\over\partial\epsilon} + {\partial\over\partial x}
\left({c\over 3 n_e \sigma_\|}\,{\partial \green\over\partial x}\right)
+ {\dot N_0\delta(\epsilon-\epsilon_0)\delta(x-x_0)\over \pi r_0^2
\epsilon_0^2}
- {\green \over t_{\rm esc}}
- \beta \, v_0 \, \delta(x-x_0) \, \green
\ ,
\label{eq4.2}
\end{equation}
and the associated radiation number and energy densities are given
by (cf. eqs.~[\ref{eq3.2}])
\begin{equation}
\ngreen(x) \equiv \int_0^\infty \epsilon^2 \,
\green(x_0,x,\epsilon_0,\epsilon) \, d\epsilon \ , \ \ \ \ \ 
\ugreen(x) \equiv \int_0^\infty \epsilon^3 \,
\green(x_0,x,\epsilon_0,\epsilon) \, d\epsilon
\ .
\label{eq4.3}
\end{equation}
Further simplification of the mathematical derivation is possible
if we work in terms of the new spatial variable $y$, defined by
\begin{equation}
y(x) \equiv \left(7 \over 3\right)^{-1 + x/x_{\rm st}}
\ .
\label{eq4.4}
\end{equation}
where $x_{\rm st}$ is given by equation~(\ref{eq2.20}). Note that $y \to
0$ in the far upstream region ($x \to -\infty$), and $y \to 1$ at the
surface of the star ($x \to x_{\rm st}$).

Based on equations~(\ref{eq2.19}) and (\ref{eq4.4}), we find that the
variation of the velocity $v$ as a function of the new variable $y$ is
given by the simple expression
\begin{equation}
v(y) = {7 \, v_c \over 4} \, (1-y)
\ .
\label{eq4.5}
\end{equation}
Likewise, we can also combine equations~(\ref{eq2.27}) and (\ref{eq4.4})
to show that the exact dynamical solution for the radiation pressure
profile as a function of $y$ is given by
\begin{equation}
P(y) = {7 \over 4} \, J \, v_c \, y
\ .
\label{eq4.6}
\end{equation}
Note that in the limit $y \to 1$, the pressure approaches the stagnation
value $P_{\rm st} \equiv (7/4) J v_c$ in agreement with equation~(\ref{eq2.28}).
Equations~(\ref{eq2.27}) and (\ref{eq4.6}) will prove useful when we seek to
confirm the validity of the results obtained for the photon distribution
$f(x_0,x,\epsilon)$ in \S~5.

Utilizing equations~(\ref{eq2.4}), (\ref{eq2.12}), (\ref{eq2.16}), and
(\ref{eq4.5}) along with the differential relation
\begin{equation}
{dx \over dy} = {r_0 \over 2 \sqrt{3}} \, \left(\sigperp \over
\sigpar\right)^{1/2} y^{-1}
\ ,
\label{eq4.7}
\end{equation}
we find that equation~(\ref{eq4.2}) can be transformed from $x$ to
$y$ to obtain
\begin{eqnarray}
y \, (1-y) \, {\partial^2 \green \over \partial y^2}
&+& \left({1 - 5 \, y \over 4}\right) {\partial \green \over \partial y}
- {\epsilon \over 4} \, {\partial \green \over \partial \epsilon}
+ \left({y - 1 \over 4 \, y}\right) \green
\nonumber
\\
&=& {3 \, \beta \, v_0 \, \delta(y-y_0) \, \green \over 7 \, v_c}
- {3 \, \dot N_0 \, \delta(\epsilon-\epsilon_0) \, \delta(y-y_0)
\over 7 \, \pi \, r_0^2 \, \epsilon_0^2 \, v_c}
\ ,
\label{eq4.8}
\end{eqnarray}
where we have made the definition
\begin{equation}
y_0 \equiv y(x_0) = \left(7 \over 3\right)^{-1 + x_0/x_{\rm st}}
\ ,
\label{eq4.9}
\end{equation}
so that $y_0$ denotes the value of $y$ at the top of the thermal mound
(see eq.~[\ref{eq4.4}]). According to equation~(\ref{eq4.5}), the flow
velocity at the top of the mound, $v_0$, is related to $y_0$ via
\begin{equation}
{v_0 \over v_c} = {7 \over 4} \, (1 - y_0)
\ .
\label{eq4.10}
\end{equation}
Note that we can now write the Green's function as either $\green(x_0,x,
\epsilon_0,\epsilon)$ or $\green(y_0,y,\epsilon_0,\epsilon)$ since
$(x,x_0)$ and $(y,y_0)$ are interchangeable via equations~(\ref{eq4.4})
and (\ref{eq4.9}).

\subsection{Separation Solutions}

When $\epsilon > \epsilon_0$, equation~(\ref{eq4.8}) is separable
in energy and space using the functions
\begin{equation}
f_\lambda(\epsilon,y) = \epsilon^{-\lambda} \, g(\lambda,y) \ ,
\label{eq4.11}
\end{equation}
where $\lambda$ is the separation constant, and the spatial function
$g$ satisfies the differential equation
\begin{equation}
y \, (1-y) \, {d^2 g \over dy^2}
+ \left({1 - 5 \, y \over 4}\right) {d g \over dy}
+ \left({\lambda \, y + y - 1 \over 4 \, y}\right) g
= {3 \, \beta \, v_0 \, \delta(y-y_0) \over 7 \, v_c} \, g
\ .
\label{eq4.12}
\end{equation}
In order to avoid an infinite spatial diffusion flux at $y=y_0$,
the function $g$ must be continuous there, and consequently we
obtain the condition
\begin{equation}
\lim_{\varepsilon \to 0} \ g(\lambda,y_0+\varepsilon)
- g(\lambda,y_0-\varepsilon) \ = \ 0
\ .
\label{eq4.13}
\end{equation}
We can also derive a jump condition for the derivative $dg/dy$ at the
top of the mound by integrating equation~(\ref{eq4.12}) with respect to
$y$ in a very small region around $y=y_0$. The result obtained is
\begin{equation}
\lim_{\varepsilon \to 0} \
{d g \over d y}\Bigg|_{y=y_0+\varepsilon}
- {d g \over d y}\Bigg|_{y=y_0-\varepsilon}
= {3 \, \beta \over 4 \, y_0} \ g(\lambda,y_0)
\ ,
\label{eq4.14}
\end{equation}
where we have also used equation~(\ref{eq4.10}). The spatial
eigenfunctions for our problem are denoted by
\begin{equation}
g_n(y) \equiv g(\lambda_n,y)
\ ,
\label{eq4.15}
\end{equation}
where $\lambda_n$ represents the $n$th eigenvalue. These functions
satisfy the continuity and derivative jump conditions given by
equations~(\ref{eq4.13}) and (\ref{eq4.14}), respectively, as
well as the boundary conditions discussed below.

The homogeneous version of equation~(\ref{eq4.12}) for $g$ obtained
when $y \ne y_0$ has fundamental solutions given by
\begin{equation}
\varphi_1(\lambda,y) \equiv y \, F(a, \, b; \, c; \, y)
\ ,
\label{eq4.16}
\end{equation}
\begin{equation}
\varphi^*_1(\lambda,y) \equiv y^{-1/4} \, F(a-5/4, \, b-5/4;
\, 2-c; \, y)
\ ,
\label{eq4.17}
\end{equation}
where $F(a,b;c;z)$ denotes the hypergeometric function (Abramowitz
\& Stegun 1970), and the parameters $a$, $b$, and $c$ are defined by
\begin{equation}
a \equiv {9 - \sqrt{17 + 16 \, \lambda} \over 8}\ , \ \ \ \ \ 
b \equiv {9 + \sqrt{17 + 16 \, \lambda} \over 8}\ , \ \ \ \ \ 
c \equiv {9 \over 4}
\ .
\label{eq4.18}
\end{equation}
The ``seed'' photons injected into the flow from the thermal mound
located near the base of the column are unable to diffuse very far up
into the accreting gas due to the extremely high speed of the inflow.
Most of the photons therefore escape through the walls of the column
within a few scattering lengths of the mound, forming a ``fan'' type
beam pattern, as expected for accretion-powered X-ray pulsars (e.g.,
Harding 1994, 2003). Based on this physical picture, we conclude that the
eigenfunction $g_n$ must {\it vanish} in the upstream limit, $y \to 0$.
Conversely, in the downstream limit, $y \to 1$, the gas settles onto the
surface of the star and the radiation pressure achieves the stagnation
value given by equation~(\ref{eq2.28}). Since this is a finite pressure,
we conclude that $g_n$ must approach a constant as $y \to 1$.

Analysis of the asymptotic behaviors of the hypergeometric functions
$\varphi_1$ and $\varphi_1^*$ shows that the first function vanishes and
the second diverges as $y \to 0$. Hence in the region upstream from the
thermal mound ($y \le y_0$), the eigenfunction $g_n$ must be given by
$\varphi_1$. However, in the downstream region ($y \ge y_0$), the
situation is more complicated because the two functions $\varphi_1$ and
$\varphi^*_1$ each diverge logarithmically as $y \to 1$. We must
therefore utilize a suitable linear combination of these functions in
order to obtain a convergent solution in the downstream region.
Consequently we define the new function (see Appendix~A for details)
\begin{equation}
\varphi_2(\lambda,y) \equiv {\Gamma(b) \over \Gamma(c) \, \Gamma(1-b)}
\ \varphi_1(\lambda,y)
- {\Gamma(1-a) \over \Gamma(2-c) \, \Gamma(a)}
\ \varphi^*_1(\lambda,y) \ ,
\label{eq4.19}
\end{equation}
which remains finite in the limit $y \to 1$ as required.

\subsection{Eigenfunctions and Eigenvalues}

By utilizing the various relations derived above, we find that the global
solution for the spatial eigenfunction $g_n(y)$ (eq.~[\ref{eq4.15}])
can now be written as
\begin{equation}
g_n(y) = \cases{
\varphi_1(\lambda_n,y) \ , & $y \le y_0$ \ , \cr
B_n \, \varphi_2(\lambda_n,y) \ , & $y \ge y_0$ \ , \cr
}
\label{eq4.20}
\end{equation}
where the constant $B_n$ is determined using the continuity condition
(eq.~[\ref{eq4.13}]), which yields
\begin{equation}
B_n = {\varphi_1(\lambda_n,y_0) \over \varphi_2(\lambda_n,y_0)}
\ .
\label{eq4.21}
\end{equation}
We can combine equations~(\ref{eq4.14}), (\ref{eq4.20}), and
(\ref{eq4.21}) to show that the eigenvalue equation for $\lambda$
is given by
\begin{equation}
\varphi_1 \, {\partial \varphi_2 \over \partial y}
- \varphi_2 \, {\partial \varphi_1 \over \partial y}
- {3 \, \beta \, \varphi_1 \, \varphi_2 \over 4 \, y}
\ \Bigg|_{y=y_0} = 0
\ .
\label{eq4.22}
\end{equation}
The left-hand side of this expression can be evaluated using the
exact solution for the Wronskian (see Appendix~B),
\begin{equation}
W(\lambda,y) \equiv
\varphi_1 \, {\partial\varphi_2 \over \partial y}
- \varphi_2 \, {\partial\varphi_1 \over \partial y}
= {5 \over 4} \, {\Gamma(1-a) \over \Gamma(a) \, \Gamma(2-c)}
\ {y^{-1/4} \over 1 - y}
\ ,
\label{eq4.23}
\end{equation}
which is applicable for arbitrary values of $\lambda$ and $y$.
Equations~(\ref{eq4.22}) and (\ref{eq4.23}) can be combined to
obtain the alternative form for the eigenvalue equation,
\begin{equation}
{5 \over 3} \, {\Gamma(1-a) \over \Gamma(a) \, \Gamma(2-c)}
\, {y_0^{3/4} \over 1 - y_0} = \beta \, \varphi_1(\lambda,y_0)
\, \varphi_2(\lambda,y_0)
\ ,
\label{eq4.24}
\end{equation}
where $a$, $b$, and $c$ are functions of $\lambda$ given by
equations~(\ref{eq4.18}). The roots of this expression are the
eigenvalues $\lambda = \lambda_n$, and the associated eigenfunctions
are evaluated using equation~(\ref{eq4.20}).

The first eigenvalue, $\lambda_0$, is especially important because it
determines the slope of the high-energy spectrum emerging from the
accretion column according to equation~(\ref{eq4.11}). The spectral
index of the emitted photon count rate distribution, $\alpha_0$, is
related to $\lambda_0$ via $\alpha_0 = \lambda_0-2$. In Figure~2 we plot
the photon index $\alpha_0$ as a function of the dimensionless
parameters $\beta$ and $y_0$. Note that $\alpha_0$ is a double-valued
function of $y_0$ for fixed $\beta$, which is a consequence of the
imposed velocity profile (eq.~[\ref{eq4.5}]). Physically, this behavior
reflects the fact that it is always possible to achieve a desired amount
of compression (first-order Fermi energization) by placing the source in
a specific location in either the upstream or downstream region of the
flow. We also observe that if we increase the absorption parameter
$\beta$ while holding $y_0$ fixed, then $\alpha_0$ increases
monotonically, and therefore the high-energy spectrum becomes
progressively steeper. This behavior is expected physically because as
the absorption parameter is increased, the injected photons spend less
time on average being energized by collisions with electrons before
either escaping from the column or being absorbed at the source
location. The reduction in the amount of energization naturally leads to
a steepening of the radiation spectrum. When $\beta = 0$, no absorption
occurs, and the index $\alpha_0$ achieves it minimum (limiting) value of
2. This limit is unphysical, however, since it yields a divergent result
for the total photon energy density $\ugreen$ according to equation
(\ref{eq4.3}).

\begin{figure}[t]
\begin{center}
\epsfig{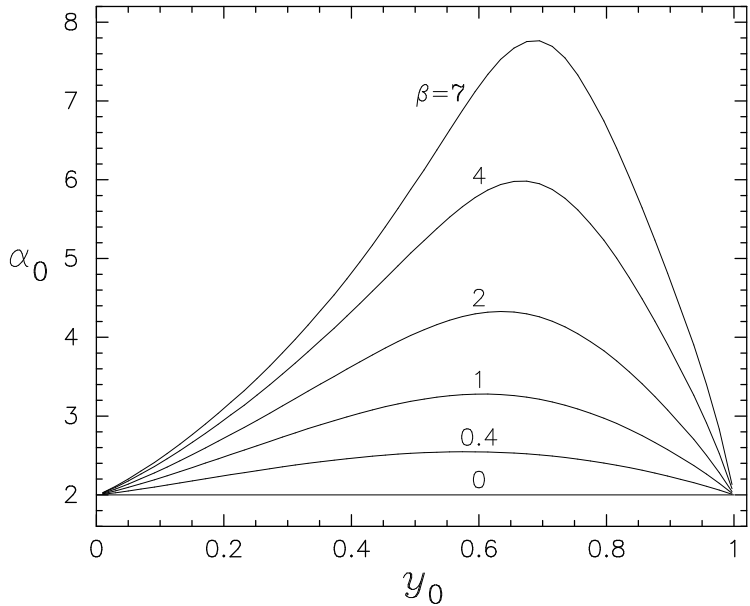}
\end{center}
\caption{High-energy power-law photon spectral index $\alpha_0$ plotted
as a function of the source location $y_0$ for the indicated values of
the absorption parameter $\beta $. Note the steepening of the radiation
spectrum that occurs when $\beta$ is increased for a fixed value of
$y_0$, which reflects the decreasing residence time for the photons in
the plasma due to the enhanced absorption.}
\end{figure}

\subsection{Eigenfunction Expansion}

We demonstrate in Appendix~C that the eigenfunctions $g_n(y)$ form an
orthogonal set, as expected since this is a standard Sturm-Liouville
problem. Once the eigenvalues and eigenfunctions are determined, the
solution for the Green's function can therefore be expressed as the
infinite series
\begin{equation}
\green(y_0,y,\epsilon_0,\epsilon)
= \epsilon^{-3} \sum_{n=0}^\infty \ C_n
\left(\epsilon \over\epsilon_0\right)^{3-\lambda_n} g_n(y)
\ ,
\label{eq4.25}
\end{equation}
where the expansion coefficients $C_n$ are computed by employing the
orthogonality of the eigenfunctions along with the condition
\begin{equation}
\green(y_0,y,\epsilon_0,\epsilon)\bigg|_{\epsilon=\epsilon_0}
= {12 \, \dot N_0 \over 7 \, \pi \, r_0^2 \, \epsilon_0^3 \, v_c}
\ \delta(y - y_0)
\ ,
\label{eq4.26}
\end{equation}
which is obtained by integrating the transport equation~(\ref{eq4.8})
with respect to $\epsilon$ in a very small range surrounding the injection
energy $\epsilon_0$. The result obtained for the $n$th expansion
coefficient is
\begin{equation}
C_n = {12 \, \dot N_0 \, y_0^{-3/4} \, g_n(y_0) \over
7 \, \pi \, r_0^2 \, v_c \, I_n}
\ ,
\label{eq4.27}
\end{equation}
where the quadratic normalization integrals $I_n$ are defined by
\begin{equation}
I_n \equiv \int_0^1 y^{-3/4} \, g_n^2(y) \, dy
\ .
\label{eq4.28}
\end{equation}
In Appendix~D, we show that the normalization integrals can be evaluated
using the closed-form expression
\begin{equation}
I_n = K(\lambda_n,y_0)
\ ,
\label{eq4.29}
\end{equation}
where
\begin{equation}
K(\lambda,y) \equiv
3 \, \beta \, y^{-3/4} (1-y) \, \varphi_1^2(\lambda,y)
\left[{\Psi(a) + \Psi(1-a) \over (17 + 16 \, \lambda)^{1/2}}
- {\partial\ln\varphi_1\over\partial\lambda}
- {\partial\ln\varphi_2\over\partial\lambda}
\right]
\ ,
\label{eq4.30}
\end{equation}
and
\begin{equation}
\Psi(z) \equiv {1 \over \Gamma(z)} \, {d \Gamma(z) \over dz}
\ .
\label{eq4.31}
\end{equation}
This provides an extremely efficient alternative to numerical
integration for the computation of $I_n$. The eigenfunction expansion
converges fairly rapidly and therefore we are able to obtain an accuracy
of at least five significant figures in our calculations of $\green$ by
terminating the series in equation~(\ref{eq4.25}) after the first 20
terms.

\subsection{Green's Function for the Escaping Photon Spectrum}

Equation~(\ref{eq4.25}) represents the exact solution for the Green's
function describing the radiation spectrum inside a pulsar accretion
column resulting from the injection of $\dot N_0$ seed photons per unit
time from a monochromatic source located at $y=y_0$ (or, equivalently,
at $x=x_0$). In the escape probability approach employed here, the
associated Green's function for the {\it photon spectrum} emitted through
the walls of the cylindrical column is defined by
\begin{equation}
\greenphoton(x_0,x,\epsilon_0,\epsilon) \equiv {\pi \, r_0^2 \,
\epsilon^2 \over t_{\rm esc}} \, \green(x_0,x,\epsilon_0,\epsilon)
\ ,
\label{eq4.32}
\end{equation}
so that $\greenphoton \, dx \, d\epsilon$ represents the number of
photons escaping from the column per unit time between positions $x$ and
$x + dx$ with energy between $\epsilon$ and $\epsilon + d\epsilon$. We
remind the reader that the quantities $(x,x_0)$ and $(y,y_0)$ are
interchangeable via equations~(\ref{eq4.4}) and (\ref{eq4.9}), and
therefore we are free to work in terms of the more convenient parameters
$(y,y_0)$ without loss of generality. By substituting for $t_{\rm esc}$
using equation~(\ref{eq2.12}) we can obtain the alternative expression
\begin{equation}
\greenphoton(y_0,y,\epsilon_0,\epsilon) = {\pi \, m_p
\, c \, v(y) \, \epsilon^2\over J \, \sigperp}
\green(y_0,y,\epsilon_0,\epsilon)
\ .
\label{eq4.33}
\end{equation}
Eliminating $J$ using equation~(\ref{eq2.16}) and substituting for $v$
and $\green$ using equations~(\ref{eq4.5}) and (\ref{eq4.25}),
respectively, yields the equivalent result
\begin{equation}
\greenphoton(y_0,y,\epsilon_0,\epsilon)
= (1-y) \, \epsilon^{-1} \sum_{n=0}^\infty \ D_n
\left(\epsilon \over\epsilon_0\right)^{3-\lambda_n} g_n(y)
\ ,
\label{eq4.34}
\end{equation}
where the expansion coefficients $D_n$ are defined by
\begin{equation}
D_n \equiv {2 \sqrt{3} \, \dot N_0 \, g_n(y_0) \over
r_0 \, y_0^{3/4} I_n}
\left({\sigpar \over \sigperp}\right)^{1/2}
\ .
\label{eq4.35}
\end{equation}

The Green's function $\greenphoton$ describes the photon distribution
escaping from the accretion column as a function of energy $\epsilon$
and location $y$ for the case of monoenergetic photon injection.
Analysis of $\greenphoton$ therefore reveals some interesting details
about the energization of the photons as they are transported through
the column via diffusion and advection, and ultimately escape through
the column walls. We plot $\greenphoton$ as a function of the energy
ratio $\epsilon/\epsilon_0$ and the location $y$ in Figure~3 for the
parameter values $\beta = 0.4$ and $y_0 = 0.9$. In this case the first
eigenvalue is given by $\lambda_0 = 4.231$, and the corresponding photon
index is $\alpha_0 = 2.231$ (see Fig.~2). The selected value of $y_0$
corresponds to a source located near the bottom of the accretion column,
just above the stellar surface. At the source location, $y=y_0=0.9$, the
energy spectrum extends down to the injection energy, $\epsilon_0$.
However, at all other radii the spectrum displays a steep turnover above
that energy because all of the photons at these locations have
experienced Fermi energization due to collisions with the infalling
electrons. The photons with energy $\epsilon = \epsilon_0$ at the source
location have been injected so recently that they have not yet
experienced significant energization. Note that for small values of $y$,
the escaping spectrum is greatly attenuated due to the inability of the
photons to diffuse upstream through the rapidly infalling plasma. The
escaping spectrum is also attenuated in the far downstream region ($y
\sim 1$) due to the very high opacity of the flow, which inhibits the
diffusion of photons through the walls. This latter effect stems from
the divergence of the electron density near the stellar surface. The
energy of maximum brightness, corresponding to the peak in
$\greenphoton$, achieves its maximum value as $y \to 0$ because the
photons that manage to diffuse far upstream from the source are the ones
that have resided in the flow the longest and therefore experienced the
most energy amplification.

\begin{figure}[t]
\begin{center}
\epsfig{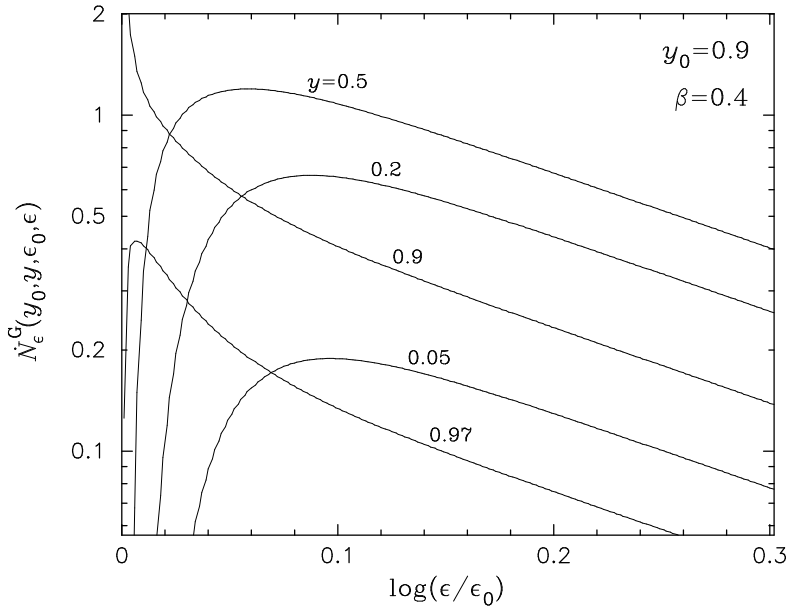}
\end{center}
\caption{Green's function $\greenphoton(y_0,y,\epsilon_0, \epsilon)$
describing the photon distribution escaping from the accretion column
per unit time (eq.~[\ref{eq4.34}]) plotted in units of $\dot
N_0(\sigpar/\sigperp)^{1/2}(r_0 \, \epsilon_0)^{-1}$ as a function of
the photon energy ratio $\epsilon/\epsilon_0$ for the indicated values
of the spatial variable $y$. In this example we have set the absorption
constant $\beta = 0.4$ and the source location $y_0 = 0.9$, so that the
monoenergetic source is located near the base of the accretion column.}
\end{figure}

We plot $\greenphoton$ as a function of energy and location
in Figure~4 for the parameters $\beta = 4$ and $y_0 = 0.4$, in which
case we obtain for the first eigenvalue and the photon index $\lambda_0
= 6.325$ and $\alpha_0 = 4.325$, respectively. In this scenario the
source is located in the upstream region and the absorption is stronger
than that in Figure~3. Due to the increased absorption resulting from
the larger value of $\beta$, the photons spend less time on average in
the flow being energized by collisions with the electrons before they
escape through the column walls or are ``recycled'' by absorption. This
causes a steepening of the spectrum at high energies, as evidenced by
the increase in the photon index $\alpha_0$. In this situation the
energy of maximum brightness (where $\greenphoton$ displays a
peak) achieves its greatest value in the downstream region. This is the
reverse of the behavior displayed in Figure~3 because in the present
example, the source is located in the upstream region and therefore the
photons that diffuse further upstream towards $y = 0$ do not experience
as much compression as those that are advected downstream. The escaping
photon distribution in the far upstream and downstream regions is
strongly attenuated due to the same processes operative in Figure~3, and
therefore most of the escaping radiation is emitted from the accretion
column around the source location.

The analytical results for the Green's function obtained in this section
provide the basis for the consideration of any source distribution since
the fundamental differential equation (\ref{eq3.1}) is linear. This is
further discussed in \S~5, where we use equations~(\ref{eq3.5}) and
(\ref{eq4.1}) to convolve the Green's function with the blackbody
spectrum produced by the thermal mound.

\begin{figure}[t]
\begin{center}
\epsfig{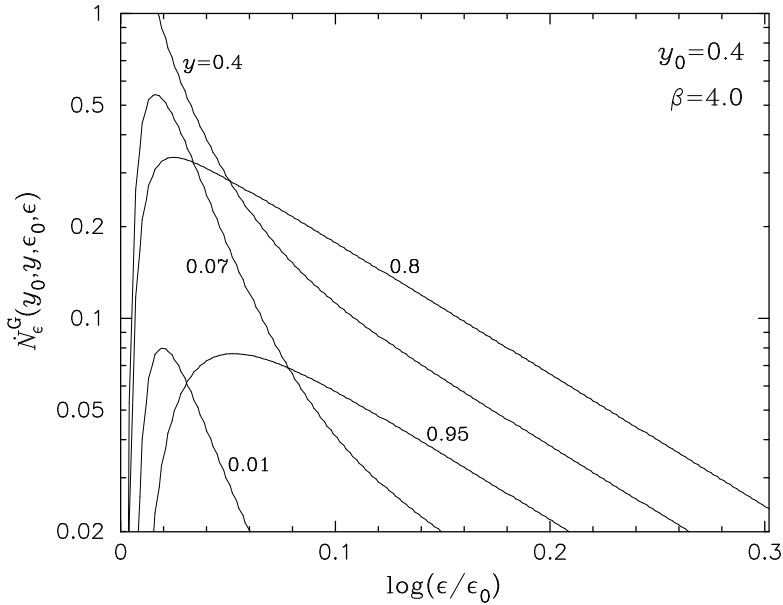}
\end{center}
\caption{Same as Fig.~3, except $\beta = 4.0$ and $y_0 = 0.4$. In this
case the source is located in the upstream region, and the absorption at
the source location is stronger than in Fig.~3. The latter effect causes
a significant steepening of the spectrum at high energies, as explained
in the text.}
\end{figure}

\subsection{Column-Integrated Green's Function}

By integrating over the vertical structure of the accretion column, we
can compute the total emitted radiation distribution, which corresponds
to the phase-averaged spectrum of the X-ray pulsar. In the cylindrical
geometry employed here, the formal integration domain is the region
$-\infty < x < x_{\rm st}$, where $x_{\rm st}$ is the distance between
the sonic point and the stellar surface (see eq.~[\ref{eq2.20}]). For
the case of a monochromatic source, we define the {\it column-integrated
Green's function} by writing
\begin{equation}
\greencolumn(x_0,\epsilon_0,\epsilon)
\equiv \int_{-\infty}^{x_{\rm st}} \greenphoton
(x_0,x,\epsilon_0,\epsilon)
\, dx
\ ,
\label{eq4.36}
\end{equation}
where $\greencolumn \, d\epsilon$ represents the number of photons
escaping from the column per unit time with energy between $\epsilon$
and $\epsilon + d\epsilon$. Using equation~(\ref{eq4.7}), the variable
of integration can be transformed from $x$ to $y$ to obtain the
alternative form
\begin{equation}
\greencolumn(y_0,\epsilon_0,\epsilon)
= {r_0 \over 2 \sqrt{3}} \left({\sigperp \over \sigpar}\right)^{1/2}
\int_0^1 \greenphoton
(y_0,y,\epsilon_0,\epsilon)
\, {dy \over y}
\ ,
\label{eq4.37}
\end{equation}
where $y_0$ is related to $x_0$ via equation~(\ref{eq4.9}). Despite the
appearance of the factor $y^{-1}$ inside the integral in
equation~(\ref{eq4.37}), the contribution from small values of $y$ is
actually negligible because the spectrum declines exponentially in the
upstream region due to advection. Furthermore, the escaping spectrum is
also strongly attenuated in the downstream region due to the divergence
of the electron density. Hence most of the radiation is emitted from the
column around $y \sim 0.5$, as indicated in Figures~3 and 4.

By combining equations~(\ref{eq4.34}), (\ref{eq4.35}), and
(\ref{eq4.37}), we can reexpress the column-integrated Green's function
as
\begin{equation}
\greencolumn(y_0,\epsilon_0,\epsilon)
= {\dot N_0 \over \epsilon \, y_0^{3/4}}
\ \sum_{n=0}^\infty \ {g_n(y_0) \over I_n}
\left(\epsilon \over\epsilon_0\right)^{3-\lambda_n}
\, X_n
\ ,
\label{eq4.38}
\end{equation}
where the quadratic normalization integrals $I_n$ are evaluated using
equation~(\ref{eq4.29}) and we have made the definition
\begin{equation}
X_n \equiv \int_0^1 g_n(y) \, (1-y) \, y^{-1} \, dy
\ .
\label{eq4.39}
\end{equation}
In Appendix~E, we demonstrate that the integral in the expression for
$X_n$ can be carried out analytically to obtain
\begin{eqnarray}
X_n &=& {\Gamma(1-a) \, (1-y_0)^2 \over \Gamma(a) \,
\varphi_2(\lambda_n,y_0)}
\bigg\{{y_0^{2-c} \, \varphi_1(\lambda_n,y_0) \over \Gamma(3-c)}
\bigg[
F(2-a,2-b;3-c;y_0)
\nonumber
\\
&+& {y_0 \over 3-c} \, F(3-a,3-b;4-c;y_0) \bigg]
+ {\varphi_1^*(\lambda_n,y_0) \over (a-1) \, (b-1) \, \Gamma(1-c)}
\bigg[
F(a,b\,;c-1;y_0)
\nonumber
\\
&+& {(c-2) \over (a-2) \, (b-2)} \, F(a,b;c-2;y_0)\bigg]
\bigg\}
+ {(1-c) \over (a-1) \, (b-1)}
\left[1 + {(c-2) \over (a-2) \, (b-2)}\right]
\ ,
\qquad
\label{eq4.40}
\end{eqnarray}
where $\varphi_1(\lambda_n,y_0)$, $\varphi_1^*(\lambda_n,y_0)$, and
$\varphi_2(\lambda_n,y_0)$ are given by equations~(\ref{eq4.16}),
(\ref{eq4.17}), and (\ref{eq4.19}), respectively, and the constants $a$,
$b$, and $c$ are computed in terms of $\lambda_n$ using
equations~(\ref{eq4.18}).

In Figure~5 we plot the dependence of the column-integrated Green's
function $\greencolumn$ on the radiation energy $\epsilon$ using the
same parameters employed in Figures~3 and 4. Note that when $\beta =
0.4$ and $y_0 = 0.9$, which corresponds to the thin line in Figure~5,
the column-integrated spectrum displays a peak at $\epsilon/\epsilon_0
\sim 6$. The peak forms because the source is located close to the
bottom of the accretion column, whereas most of the photons escape from
higher altitudes around $y \sim 0.5$ (see Fig.~3). The escaping
radiation has therefore experienced significant compression in the flow
and consequently the photon energies are generally boosted well above
the injection energy $\epsilon_0$. On the other hand, when $\beta = 4$
and $y_0 = 0.4$, which corresponds to the thick line in Figure~5, the
absorption is so strong that the spectrum is quite steep and therefore
the peak is suppressed. Consequently the column-integrated spectrum
achieves it maximum value at the injection energy in this case.

\subsection{Reprocessed Blackbody Radiation}

The closed-form solutions for the Green's function $\greenphoton$
(eq.~[\ref{eq4.34}]) and for the column-integrated Green's function
$\greencolumn$ (eq.~[\ref{eq4.38}]) provide a very efficient means for
computing the X-ray spectrum escaping through the walls of the accretion
column due to the injection of monochromatic seed photons. However, in
our physical application to X-ray pulsars, we are primarily interested
in computing the emitted spectrum resulting from the reprocessing of
{\it blackbody radiation} injected into the accretion column from the
thermal mound. In the escape probability approach utilized here,
the photon distribution emitted through the walls of the cylindrical
column at position $x$, denoted by $\photonparticular(x_0,x,\epsilon)$,
is computed using
\begin{equation}
\photonparticular(x_0,x,\epsilon) \equiv {\pi \, r_0^2 \,
\epsilon^2 \over t_{\rm esc}} \, f(x_0,x,\epsilon)
\ ,
\label{eq4.41}
\end{equation}
where the particular solution $f(x_0,x,\epsilon)$ is evaluated using
equation~(\ref{eq4.1}), and $\photonparticular \, dx \, d\epsilon$
represents the number of photons escaping from the column per unit time
between positions $x$ and $x + dx$ with energy between $\epsilon$ and
$\epsilon + d\epsilon$. By combining equations~(\ref{eq4.1}) and
(\ref{eq4.41}), we find that the escaping photon spectrum can be
expressed as
\begin{equation}
\photonparticular(x_0,x,\epsilon)
= {\pi \, r_0^2 \, \epsilon^2 \over t_{\rm esc}}
\int_0^\epsilon {\green(x_0,x,\epsilon_0,\epsilon)
\over \dot N_0} \ \epsilon_0^2 \, S(\epsilon_0) \, d\epsilon_0
\ ,
\label{eq4.42}
\end{equation}
where the blackbody source function $S(\epsilon_0)$ is evaluated using
equation~(\ref{eq3.5}). The spatial variables $(x,x_0)$ and $(y,y_0)$
are interchangeable by virtue of equations~(\ref{eq4.4}) and
(\ref{eq4.9}), and therefore we can use equations~(\ref{eq4.32}) and
(\ref{eq4.42}) to write
\begin{equation}
\photonparticular(y_0,y,\epsilon)
= \int_0^\epsilon {\greenphoton(y_0,y,\epsilon_0,\epsilon)
\over \dot N_0} \ \epsilon_0^2 \,
S(\epsilon_0) \, d\epsilon_0
\ ,
\label{eq4.43}
\end{equation}
where $\greenphoton(y_0,y,\epsilon_0,\epsilon)$ is computed using
equation~(\ref{eq4.34}).

\begin{figure}[t]
\begin{center}
\epsfig{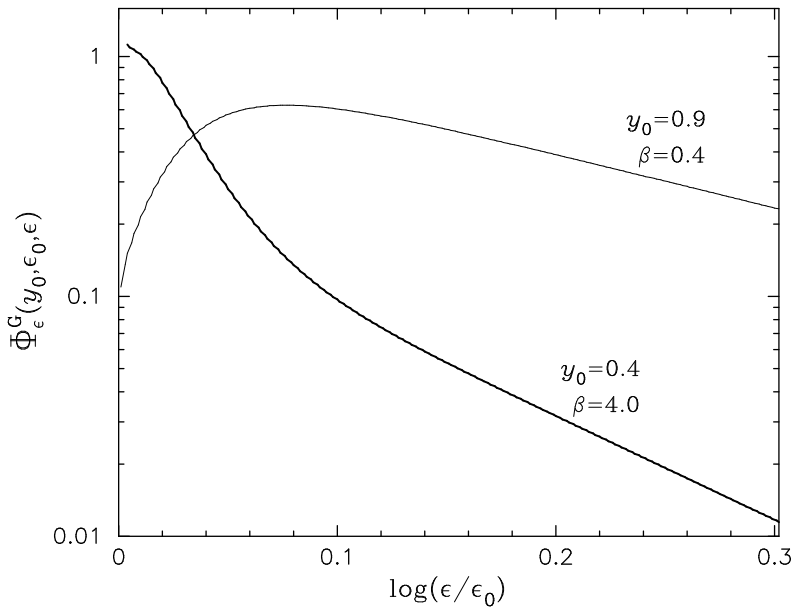}
\end{center}
\caption{Column-integrated Green's function $\greencolumn
(y_0,\epsilon_0,\epsilon)$ describing the photon spectrum escaping
through the walls of the accretion column (eq.~[\ref{eq4.38}]) plotted
in units of $\dot N_0/\epsilon_0$ as a function of the photon energy
ratio $\epsilon/\epsilon_0$. The parameter values used are $y_0 = 0.9$,
$\beta = 0.4$ ({\it thin line}) and $y_0 = 0.4$, $\beta = 4.0$ ({\it
thick line}), which correspond to the spectra plotted in Figs.~3 and 4,
respectively.}
\end{figure}

By analogy with equation~(\ref{eq4.36}) for the Green's function, we
define the particular solution for the {\it column-integrated} photon
spectrum escaping from the plasma due to the blackbody source,
$\column(x_0,\epsilon)$, by writing
\begin{equation}
\column(x_0,\epsilon)
\equiv \int_{-\infty}^{x_{\rm st}} \photonparticular(x_0,x,\epsilon)
\, dx
\ ,
\label{eq4.44}
\end{equation}
so that $\column \, d\epsilon$ gives the number of photons escaping from
the column per unit time with energy between $\epsilon$ and $\epsilon +
d\epsilon$. The variable of integration can be transformed from $x$ to
$y$ using equation~(\ref{eq4.7}), which yields
\begin{equation}
\column(y_0,\epsilon)
= {r_0 \over 2 \sqrt{3}} \left({\sigperp \over \sigpar}\right)^{1/2}
\int_0^1 \photonparticular(y_0,y,\epsilon)
\, {dy \over y}
\ ,
\label{eq4.45}
\end{equation}
where $y_0$ is related to $x_0$ by equation~(\ref{eq4.9}). By
substituting for $\photonparticular$ using equation~(\ref{eq4.43}) and
employing equation~(\ref{eq4.37}), we can obtain the alternative form
\begin{equation}
\column(y_0,\epsilon)
= \int_0^\epsilon {\greencolumn(y_0,\epsilon_0,\epsilon)
\over \dot N_0} \ \epsilon_0^2 \,
S(\epsilon_0) \, d\epsilon_0
\ .
\label{eq4.46}
\end{equation}
The column-integrated Green's function $\greencolumn
(y_0,\epsilon_0,\epsilon)$ is evaluated using equation~(\ref{eq4.38}).
In \S~5 we will utilize equations~(\ref{eq4.43}) and (\ref{eq4.46}) to
compute the photon spectrum emitted from the accretion column using
parameters corresponding to specific X-ray pulsars.

\section{ASTROPHYSICAL APPLICATIONS}

For given values of the stellar mass $M_*$ and the stellar radius $R_*$,
our model has three free parameters, namely the column radius $r_0$, the
temperature at the top of the thermal mound $T_0$, and the accretion
rate $\dot M$. In this section we investigate the relationship between
these quantities and the dimensionless parameters $y_0$ and $\beta$
appearing in the theory. Although it is not our intention in this paper
to develop complete models for X-ray pulsar spectra, it is nonetheless
interesting to compare the simplified model developed here with actual
data for a few sources. Hence in this section we will also compute the
spectrum emerging from the accretion column for parameters corresponding
to two specific X-ray pulsars.

\subsection{Location of the Thermal Mound}

In order to understand how the theory developed here can be related to
observables such as the temperature and the luminosity, we first need to
determine how the position of the thermal mound, $x_0$ (or,
equivalently, $y_0$), depends on the values of $r_0$, $T_0$, and $\dot
M$. The blackbody surface of the mound represents the photosphere for
photon creation and destruction in this problem, and therefore the
opacity is dominated by free-free absorption inside the mound. For a
given value of the temperature $T_0$, the density at the top of the
mound, $\rho_0$, can be calculated by setting the free-free optical
thickness of the column equal to unity, so that
\begin{equation}
\tau_0^{\rm ff} \equiv r_0 \, \alpha_{\rm R}^{\rm ff}(r_0) = 1
\ ,
\label{eq5.1}
\end{equation}
where the Rosseland mean of the free-free absorption coefficient is
evaluated in cgs units using (Rybicki \& Lightman 1979)
\begin{equation}
\alpha_{\rm R}^{\rm ff}(r_0) = 6.10 \times 10^{22} \ T_0^{-7/2} \ \rho_0^2
\ ,
\label{eq5.2}
\end{equation}
for pure, fully-ionized hydrogen with the Gaunt factor set equal
to unity. The density at the top of the thermal mound is therefore
given by
\begin{equation}
\rho_0 = 4.05 \times 10^{-12} \ T_0^{7/4} \ r_0^{-1/2}
\ .
\label{eq5.3}
\end{equation}
The velocity at the top of the mound, $v_0$, is related to $\rho_0$
via the continuity equation $\dot M = \pi r_0^2 \, v_0 \, \rho_0$,
so that
\begin{equation}
v_0 = 7.86 \times 10^{10} \ \dot M \, r_0^{-3/2} \, T_0^{-7/4}
\ .
\label{eq5.4}
\end{equation}
By combining this result with equations~(\ref{eq2.23}) and (\ref{eq4.10}),
we find that the value of $y_0$ is given by
\begin{equation}
y_0 = 1 - 2.15 \times 10^{14} \ R_*^{1/2} \, M_*^{-1/2} \,
\dot M \, r_0^{-3/2} \, T_0^{-7/4}
\ ,
\label{eq5.5}
\end{equation}
or, equivalently,
\begin{equation}
y_0 = 1 - 8.57 \times 10^{-4} \ \left(R_* \over 10\,\rm km\right)^{1/2}
\left(M_* \over \msun\right)^{-1/2}
\left(\!\dot M \over 10^{16}\,\rm g\,s^{-1}\!\right)
\left(r_0 \over 1\,\rm km\right)^{-3/2}
\left(T_0 \over 10^7\,\rm K\right)^{-7/4}
\ .
\label{eq5.6}
\end{equation}

The values obtained for $y_0$ are extremely close to unity, indicating
that the top of the thermal mound is just above the stellar surface.
Once $y_0$ is computed, we can determine $x_0$ by using
equation~(\ref{eq4.9}) to write
\begin{equation}
{x_0  \over x_{\rm st}} = 1 + {\ln y_0 \over \ln(7/3)}
\ ,
\label{eq5.7}
\end{equation}
where $x_{\rm st}$ is the distance between the sonic point and the
stellar surface given by equation~(\ref{eq2.20}). The height of
the thermal mound above the stellar surface is then given by
(see eq.~[\ref{eq2.21}])
\begin{equation}
h_0 \equiv h(x_0) = {\ln(1/y_0) \over \ln(7/3)}
\ x_{\rm st}
\ ,
\label{eq5.8}
\end{equation}
or, equivalently,
\begin{equation}
h_0 = {r_0 \over 2 \sqrt{3}} \left(\sigperp \over \sigpar
\right)^{1/2} \ln\left(1 \over y_0\right)
\ ,
\label{eq5.9}
\end{equation}
where we have substituted for $x_{\rm st}$ using
equation~(\ref{eq2.20}). Specific numerical examples are given in
\S~5.4. For typical X-ray pulsar parameters, we find that the mound is
situated very close to the surface of the star, as expected.

\subsection{Computation of the Eigenvalues}

In order to compute the eigenvalues $\lambda_n$ using
equation~(\ref{eq4.24}), we must also evaluate the dimensionless
absorption parameter $\beta$ appearing in the transport
equation~(\ref{eq3.1}). Based on energy conservation considerations in
the radiation-dominated plasma, we have shown in \S~3.2 that the photon
energy flux $Q$ must be continuous across the location of the thermal
mound. This leads to equation~(\ref{eq3.13}), which can be rewritten as
\begin{equation}
\beta = {\sigma \, T_0^4 \over 3 \, v_0 \, P_0}
\ ,
\label{eq5.10}
\end{equation}
where $P_0 = U_0/3$ is the radiation pressure at the top of the thermal
mound. According to equation~(\ref{eq4.6}), $P_0 = (7/4) \, J \, v_c \,
y_0$, and therefore
\begin{equation}
\beta = {4 \, \sigma \, T_0^4 \over 21 \, J \, v_c \, v_0 \, y_0}
\ .
\label{eq5.11}
\end{equation}
By substituting for $v_c$ and $v_0$ using equations~(\ref{eq2.23}) and
(\ref{eq4.10}), respectively, and setting $J = \dot M/(\pi r_0^2)$,
we can obtain the alternative form
\begin{equation}
\beta = {\pi \, r_0^2 \, \sigma \, T_0^4 \, R_* \over
6 \, y_0 \, (1-y_0) \, G \, M_* \, \dot M}
\ ,
\label{eq5.12}
\end{equation}
or, equivalently,
\begin{equation}
\beta = {2.24 \times 10^{-3} \over y_0 \, (1-y_0)} \
\left(R_* \over 10\,\rm km\right)
\left(M_* \over \msun\right)^{-1}
\left(\!\dot M \over 10^{16}\,\rm g\,s^{-1}\!\right)^{-1}
\left(r_0 \over 1\,\rm km\right)^2
\left(T_0 \over 10^7\,\rm K\right)^4
\ .
\label{eq5.13}
\end{equation}
Taken together, equations~(\ref{eq5.6}) and (\ref{eq5.13}) allow the
determination of the two dimensionless model parameters $y_0$ and
$\beta$ in terms of $R_*$, $M_*$, $r_0$, $T_0$, and $\dot M$. This
closes the system and facilitates the calculation of the emergent photon
number spectrum $\photonparticular$ using equation~(\ref{eq4.43}), and the
calculation of the column-integrated spectrum $\column$
using equation~(\ref{eq4.46}). The values of the parameters $r_0$,
$T_0$, and $\dot M$ can therefore be extracted by comparing the
theoretically predicted photon count-rate distributions with X-ray
pulsar spectral data.

\subsection{Similarity Variables}

The eigenvalues $\lambda_n$ are functions of the dimensionless
parameters $y_0$ and $\beta$. Based on the dependences of these
parameters on $r_0$, $T_0$, and $\dot M$ expressed by
equations~(\ref{eq5.6}) and (\ref{eq5.13}), we note that the
eigenspectrum {\it remains unchanged} if $r_0$, $T_0$, and $\dot M$ are
simultaneously varied according to the prescription
\begin{equation}
r_0 \ \propto \ T_0^{-9/2} \ \propto \ \dot M^{9/10}
\ .
\label{eq5.14}
\end{equation}
In this case, the column-integrated Green's function (eq.~[\ref{eq4.38}])
also remains unchanged since it depends only on the values of $\beta$
and $y_0$. These findings suggest that it is useful to introduce the new
``similarity variables'' $p$ and $q$, defined by
\begin{equation}
p \equiv
\left(T_0 \over 10^7\,\rm K\right)
\left(\!\dot M \over 10^{16} {\rm \,g\,s}^{-1}\!\right)^{1/5}
\ , \ \ \ \ \
q \equiv 
\left(T_0 \over 10^7\,\rm K\right)
\left(r_0 \over 1\,\rm km\right)^{2/9}
\ .
\label{eq5.15}
\end{equation}
In terms of these parameters, equations~(\ref{eq5.6}) and (\ref{eq5.13})
for $y_0$ and $\beta$ now become
\begin{equation}
y_0 = 1 - 8.57 \times 10^{-4}
\left(R_* \over 10\,\rm km\right)^{1/2}
\left(M_* \over \msun\right)^{-1/2}
p^5 \, q^{-27/4}
\ ,
\label{eq5.16}
\end{equation}
and
\begin{equation}
\beta = {2.24 \times 10^{-3} \over y_0 \, (1-y_0)}
\, \left(R_* \over 10\,\rm km\right)
\left(M_* \over \msun\right)^{-1}
p^{-5} \, q^9
\ .
\label{eq5.17}
\end{equation}
The introduction of the new variables $p$ and $q$ is convenient because
it reduces the dimensionality of the parameter space that determines the
eigenvalues $\lambda_n$ from the three-dimensional domain $(r_0, T_0,
\dot M)$ to the two-dimensional space $(p, q)$. In Figure~6 we plot the
spectral index $\alpha_0 = \lambda_0-2$ of the photon count-rate
spectrum as a function of $p$ and $q$, where $\lambda_0$ is the first
eigenvalue and we have set $R_* = 10\,$km and $M_* = 1.4\,\msun$ (the
same quantity was plotted as a function of $y_0$ and $\beta$ in
Figure~2). Note that when $q$ is increased for a fixed value of $p$, the
values of $y_0$ and $\beta$ both increase according to
equations~(\ref{eq5.16}) and (\ref{eq5.17}). The source therefore moves
further downstream, and the absorption at the source location becomes
stronger. This causes the spectrum to steepen as we move upwards along a
vertical line in Figure~6. However, if $q$ becomes very large, the
spectrum starts to harden again due to the increasing amount of
compression experienced as the source approaches the base of the
accretion column.

\begin{figure}[t]
\begin{center}
\epsfig{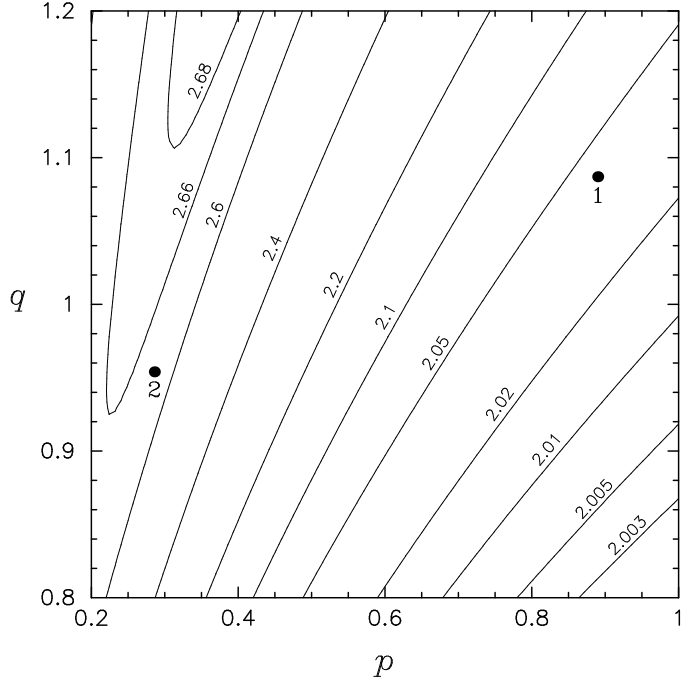}
\end{center}
\caption{Contour plot of the high-energy power-law photon spectral index
$\alpha_0$ as a function of the similarity variables $p$ and $q$, which
are related to the fundamental physical parameters $r_0$, $T_0$, and
$\dot M$ via eqs.~(\ref{eq5.15}). The value of $\alpha_0$ is indicated
for each curve, and we have set $R_* = 10\,$km and $M_* = 1.4\,\msun$.
The parameters used for models~1 and 2 are represented by the labeled
points.}
\end{figure}


\begin{figure}[t]
\begin{center}
\epsfig{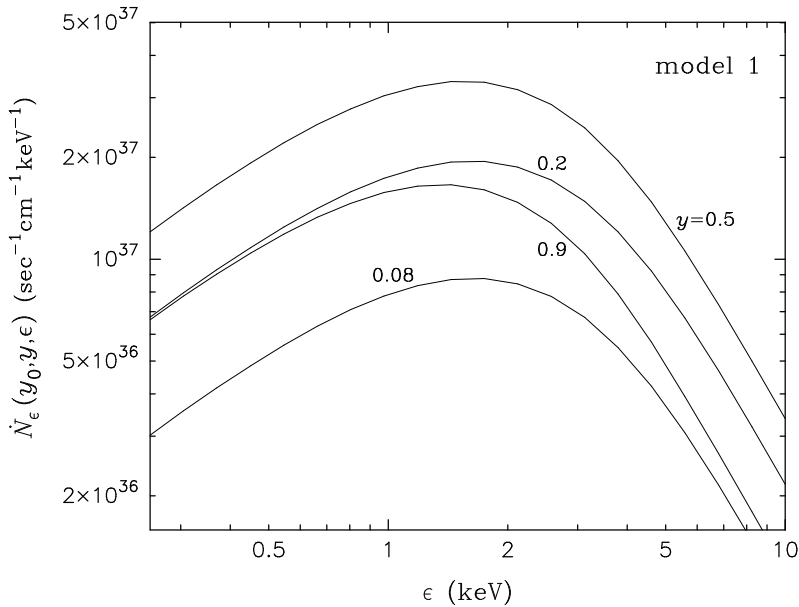}
\end{center}
\caption{Escaping photon distribution
$\photonparticular(y_0,y,\epsilon)$ (eq.~[\ref{eq4.43}]) divided by
$(\sigpar / \sigperp)^{1/2}$ plotted as a function of $\epsilon$ for the
indicated values of $y$. These results were computed using the model~1
parameters and describe the spectrum emitted per unit length of the
accretion column due to the bulk Comptonization of blackbody radiation.}
\end{figure}

\begin{figure}[t]
\begin{center}
\epsfig{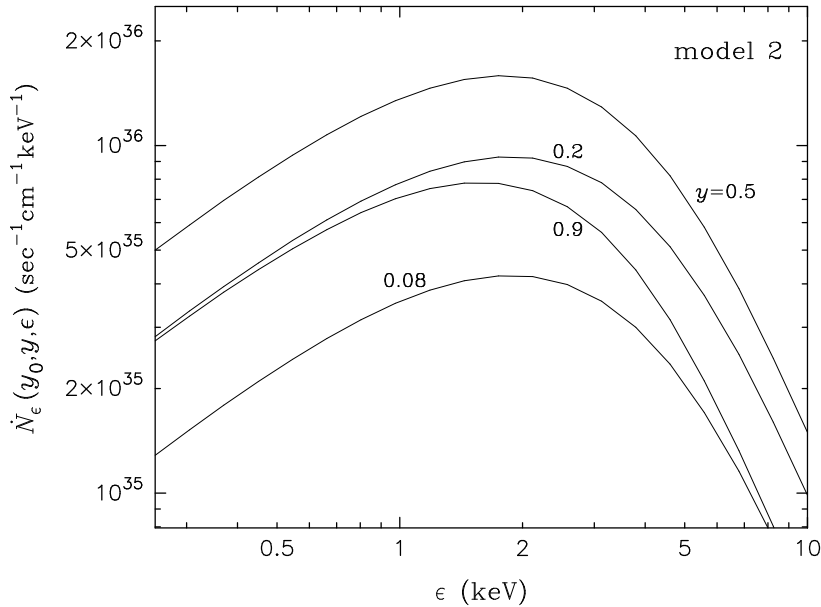}
\end{center}
\caption{Same as Fig.~7, except the results were computed using the
model~2 parameters.}
\end{figure}

\subsection{Example Spectra}

Our primary goal in this paper is to explore the effect of bulk
Comptonization on the shape of the X-ray continuum spectrum emerging
from a pulsar accretion column. The development of complete models that
include additional effects such as thermal Comptonization, cyclotron
processes, free-free emission and absorption, and iron line formation
will be pursued in future work. Here we compare spectra computed using
our simplified model with the observations for a few sources in order to
focus attention on the role of the accretion shock in the formation of
the X-ray continuum. In Figures~7 and 8 we plot the position-dependent
photon count rate spectra $\photonparticular$ (eq.~[\ref{eq4.43}])
escaping from the accretion column computed using the two sets of model
parameters listed in Table~1. These results were obtained using 20
eigenvalues and eigenfunctions, which yields extremely high precision
due to the rapid convergence of the expansion. The theoretical spectra
describe the upscattering of Planckian seed photons injected into the
base of the accretion column from the surface of the thermal mound. In
both cases we use $M_* = 1.4 \, \Msun$ and $R_* = 10\,$km for the
stellar mass and radius, respectively. For model~1 (Fig.~7), the
parameters adopted are $r_0 = 6\,$km, $T_0 = 7.3 \times 10^6\,K$, and
$\dot M = 2.69 \times 10^{16}\,{\rm g \, s}^{-1}$. In model~2 (Fig.~8)
we set $r_0 = 1.3\,$km, $T_0 = 9.0 \times 10^6\,$K, and $\dot M = 3.23
\times 10^{13}\,{\rm g \, s}^{-1}$. We use equations~(\ref{eq5.15}),
(\ref{eq5.16}), and (\ref{eq5.17}) to compute the theoretical parameters
$p$, $q$, $y_0$ and $\beta$ based on the selected values for $r_0$,
$T_0$, and $\dot M$. In model~1 we obtain $\beta = 26.4505$ and $y_0 =
0.99977$, and in model~2 we find that $\beta = 2.894 \times 10^{5}$ and
$y_0 = 0.999998$. Additional parameters are provided in Table~1. The
fact that the thermal mound location $y_0$ is extremely close to unity
in both cases indicates that the mound lies just above the surface of
the star, as expected. It is interesting to compute the actual physical
distance between the thermal mound and the stellar surface, $h_0$, which
is given by equation~(\ref{eq5.9}). Using the parameters corresponding
to models~1 and 2 yields $h_0 = 39.84\, (\sigperp/\sigpar)^{1/2}\,$cm
and $h_0 = 0.075\,(\sigperp/ \sigpar)^{1/2}\,$cm, respectively. Note
that in model~2 the thermal mound is essentially on the surface of the
star, whereas in model~1 there is a much larger separation. This
reflects the fact that the accretion rate in model~1 is over 800 times
larger than in model~2.

The results obtained for the first eigenvalue and for the photon index
in model~1 are $\lambda_0 = 4.0398$ and $\alpha_0 = 2.0398$,
respectively. In model~2, we find that the spectrum is significantly
steeper, with $\lambda_0 = 4.6382$ and $\alpha_0 = 2.6382$. The steeper
spectrum obtained in model~2 results from the large value obtained for
$\beta$, as indicated in Table~1. In general, the escaping spectrum has
a power-law shape at high energies, as expected for a Fermi process, and
a low-energy turnover due to the Planck distribution of the seed
photons. Note that the emitted radiation is concentrated in a layer just
above the thermal mound location, around $y \sim 0.5$. This reflects the
fact that advection due to collisions with the infalling stream of
high-speed electrons tends to keep the photons trapped at low altitudes
in the column, and therefore few of them are able to escape through the
column walls for $y \ll 1$. Since the spin axis and the column axis are
not aligned in pulsars, the portion of the accretion column visible
along the line of sight to the Earth changes as the pulsar rotates. When
combined with the vertical variation of the energy dependence of the
escaping photon distribution, this effect will produce a pulse-phase
dependence in the observed X-ray spectrum.

In Figures~9 and 10 we plot the photon count rate flux measured at Earth,
\begin{equation}
F_\epsilon(\epsilon) \equiv {\column(y_0,\epsilon) \over 4 \pi D^2}
\ ,
\label{eq5.18}
\end{equation}
computed using models~1 and 2, respectively, where $D$ is the distance
to the source and the column-integrated spectrum $\column(y_0,\epsilon)$
is evaluated using equation~(\ref{eq4.46}). Also included in Figures~9
and 10 for comparison are plots of the unfolded phase-averaged spectra
for the X-ray pulsars 4U~1258-61 (GX~304-1) and 4U~0352+30 (X~Persei),
respectively. The 4U~1258-61 data was reported by White et al. (1983),
and the 4U~0352+30 data is the result of XSPEC analysis of an archival
RXTE observation taken in July 1998 and reported by Delgado-Mart\'i et
al. (2001). Based on estimates from Negueruela (1998), the values
adopted for the distance $D$ are 2.5\,kpc for 4U~1258-61 and 0.35\,kpc
for 4U~0352+30. The observational data represent the de-convolved
(incident) spectra, which depend weakly on the detector response model.
The theoretical spectra were computed using 20 eigenvalues and
eigenfunctions for high accuracy, and various amounts of interstellar
absorption were included as indicated in the figures.

Although the results presented here are not fits to the data, we note
that the general shape of the pulsar spectrum predicted by the theory
agrees fairly well with the observations in each case, including both
the turnover at low energies and the power law at higher energies.
Several other sources yield similar agreement. In our model, the
turnover at $\sim 2\,$keV is due to Planckian photons that escape from
the accretion column without experiencing many scatterings. This effect
will tend to reduce the amount of absorption required to fit the
observational data, compared with the amount required using the standard
ad hoc models usually employed in X-ray pulsar spectral analysis (see
the discussion in Hickox, Narayan, \& Kallman 2004).

The second source, X~Persei, presents an interesting test for the model
due to its relatively low luminosity, $L_{\rm X} \sim 10^{34}{\rm \ ergs
\ s}^{-1}$. The agreement between the theory and the observations
suggests that radiation pressure may still be playing an important role
in the dynamics of this source, despite its low luminosity. In fact,
Langer \& Rappaport (1982) pointed out that radiation pressure may have
a strong effect on the flow dynamics of low-luminosity pulsars if the
photon energies are close to the cyclotron resonance, which greatly
increases the electron scattering cross section. Furthermore, the
dynamical importance of the gas pressure in this luminosity range may
lead to the development of a gas-mediated ``subshock'' within the
overall gradual compression of the radiation-dominated shock. This type
of hybrid shock is analogous to the ``two-fluid'' model of cosmic-ray
mediated supernova shocks (e.g., Drury \& V\"olk 1981). It is know that
in the cosmic-ray case, multiple flow solutions can occur for the same
set of upstream boundary conditions (e.g., Becker \& Kazanas 2001). There
is some possibility that the same type of multi-state behavior may
occur in X-ray pulsars, which could have interesting observational
implications.

\begin{figure}[t]
\begin{center}
\epsfig{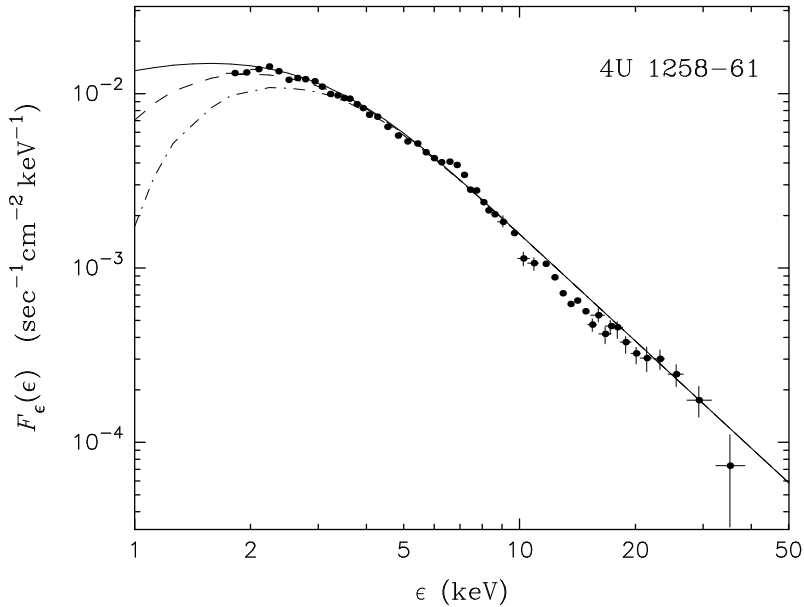}
\end{center}
\caption{Theoretical column-integrated count-rate spectrum
$F_\epsilon(\epsilon)$ (eq.~[\ref{eq5.18}]) computed using the model~1
parameters along with various amounts of interstellar absorption is
compared with the X-ray spectrum of 4U~1258-61. The column densities
used are $N_{\rm H}=0$ ({\it solid line}); $N_{\rm H}=3 \times
10^{21}\,{\rm cm}^{-2}$ ({\it dashed line}); and $N_{\rm H}=9 \times
10^{21}\,{\rm cm}^{-2}$ ({\it dot-dashed line}). The other parameters
for the theoretical model are given in Table~1.}
\end{figure}

\begin{figure}[t]
\begin{center}
\epsfig{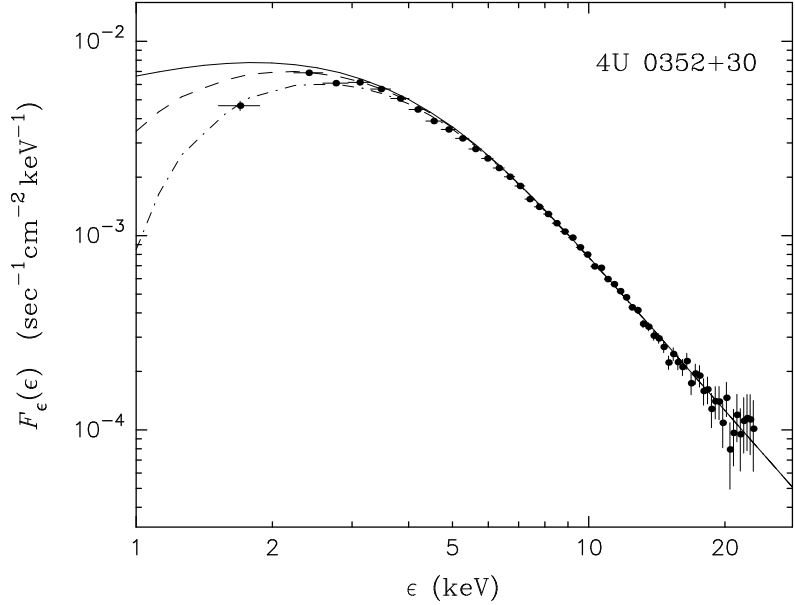}
\end{center}
\caption{Same as Fig.~9, except the data corresponds to 4U~0352+30
and the theoretical spectra were computed based on model~2. The column
densities used are $N_{\rm H}=0$ ({\it solid line}); $N_{\rm H}=3 \times
10^{21}\,{\rm cm}^{-2}$ ({\it dashed line}); and $N_{\rm H}=9 \times
10^{21}\,{\rm cm}^{-2}$ ({\it dot-dashed line}).}
\end{figure}

\section{MODEL SELF-CONSISTENCY}

The analytical approach employed in this paper is based on the
conservation of energy and momentum in the radiation-dominated flow.
This leads to the self-consistent determination of the thermal mound
location, $x_0$, for given values of the fundamental parameters $r_0$,
$T_0$, and $\dot M$, as discussed in \S~5.1. In this section we verify
that the model satisfies the self-consistency conditions required for
global energy conservation. We also examine the implications of the
one-dimensional escape probability formalism that has been utilized
to model the emission of radiation from the column.

\subsection{Pressure Distribution}

The exact dynamical solution for the radiation pressure inside the pulsar
accretion column is given by equation~(\ref{eq2.27}), which states
that
\begin{equation}
P(x) = {7 \over 4} \, J \, v_c \left(7 \over 3\right)
^{-1+x/x_{\rm st}}
\ ,
\label{eq5.19}
\end{equation}
where $x_{\rm st}$ is the distance between the sonic point and the
stellar surface, given by equation~(\ref{eq2.20}). In principle, the
radiation pressure can also be computed by integrating the particular
solution for the photon distribution $f$ inside the accretion column.
Since $P=U/3$, we can use equation~(\ref{eq3.2}) to write
\begin{equation}
P(x) = {1 \over 3} \int_0^\infty \epsilon^3 \,
f(x_0,x,\epsilon) \, d\epsilon
\ ,
\label{eq5.20}
\end{equation}
where $f$ is computed using the convolution (see eq.~[\ref{eq4.1}])
\begin{equation}
f(x_0,x,\epsilon) = \int_0^\epsilon {\green(x_0,x,\epsilon_0,\epsilon)
\over \dot N_0} \ \epsilon_0^2 \, S(\epsilon_0) \, d\epsilon_0
\ .
\label{eq5.21}
\end{equation}
Here, the blackbody source $S(\epsilon_0)$ is given by
equation~(\ref{eq3.5}) and the Green's function $\green$ is evaluated
using equation~(\ref{eq4.25}), bearing in mind that the quantities
$(x,x_0)$ and $(y,y_0)$ are interchangeable via equations~(\ref{eq4.4})
and (\ref{eq4.9}).

We can check the accuracy of the entire solution technique, including
the determination of the eigenfunctions and the eigenvalues, by
comparing the exact dynamical result for $P$ with that obtained by
integrating the spectrum. In Figure~11 we plot the exact analytical
solution for the radiation pressure computed using
equation~(\ref{eq5.19}) along with the results obtained by integrating
the model~1 and 2 spectra using equation~(\ref{eq5.20}). Note that the
agreement between the various solutions for $P$ is excellent, and in
each case the pressure achieves the correct stagnation value $P_{\rm st}
= (7/4) \, J \, v_c$ (eq.~[\ref{eq2.28}]) as the gas approaches the
stellar surface. The agreement between the various results for the
pressure confirms that our solution for the radiation spectrum is
consistent with the dynamics of the accretion flow.

\begin{figure}[t]
\begin{center}
\epsfig{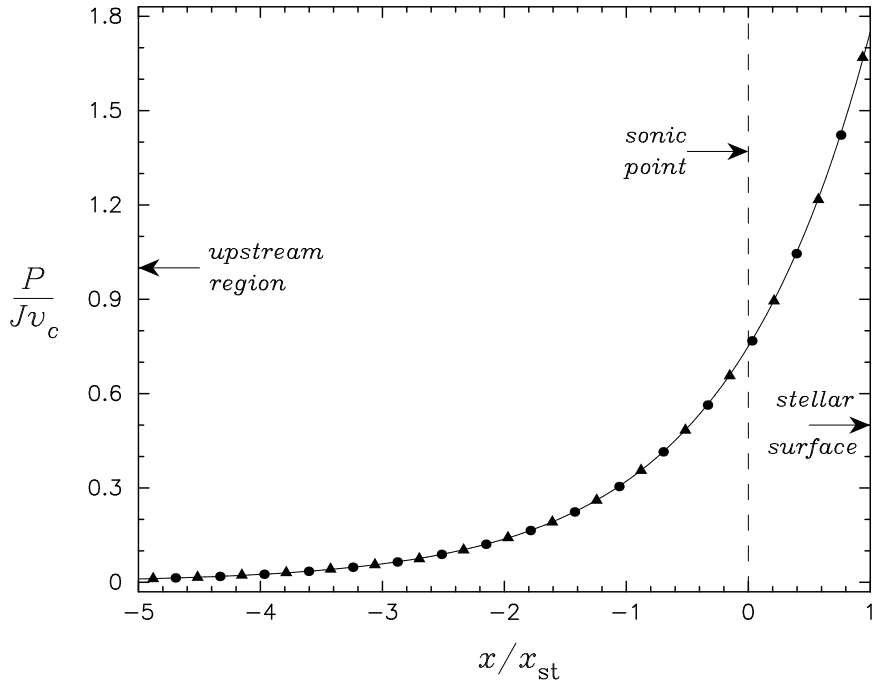}
\end{center}
\caption{Radiation pressure $P$ plotted in units of $J v_c$ as a
function of $x$, which is the distance measured from the sonic point in
the direction of the flow. The dashed vertical line at $x=0$ represents
the sonic point, and $x=x_{\rm st}$ at the stellar surface. The exact
dynamical solution (eq.~[\ref{eq5.19}]) is denoted by the solid line,
and the filled circles and triangles represent the values obtained using
eq.~(\ref{eq5.20}) to integrate the model~1 and 2 spectra, respectively.
At the surface of the star, the pressure approaches the stagnation value
(eq.~[\ref{eq2.28}]) as expected.}
\end{figure}

\subsection{Total Luminosity}

In order to further explore the self-consistency of the formalism
developed here, we have also computed the total X-ray luminosity,
$L_{\rm X}$, emerging from the column by integrating the escaping
number spectrum $\column$ (eq.~[\ref{eq4.44}]) using
\begin{equation}
L_{\rm X} = \int_0^\infty \epsilon \, \column(x_0,\epsilon)
\, d\epsilon
\ .
\label{eq5.22}
\end{equation}
We have confirmed that the values obtained using this expression
agree precisely with the accretion luminosity given by
\begin{equation}
L_{\rm X} = {G M_* \dot M \over R_*}
\ .
\label{eq5.23}
\end{equation}
The calculations performed in this section and in \S~6.1 validate our
entire solution procedure for the determination of the radiation
spectrum by demonstrating that our results are fully consistent with
global energy conservation.

\subsection{Escape-Probability Formalism}

Within the context of our one-dimensional model, the diffusion of
radiation through the walls of the accretion column is treated in an
approximate manner using an escape-probability formalism. In this
approach, the probability per unit time that a randomly-selected photon
at location $x$ inside the column will escape through the walls is equal
to $t_{\rm esc}^{-1}$, where the mean escape time $t_{\rm esc}$ is a
function of $x$ given by (see eqs.~[\ref{eq2.11}])
\begin{equation}
t_{\rm esc}(x) = {r_0^2 \, \sigperp n_e(x) \over c}
\ ,
\label{eq5.24}
\end{equation}
with $r_0$ and $\sigperp$ denoting the radius of the cylindrical column
and the electron scattering cross section for photons propagating
perpendicular to the column axis, respectively. The escape of radiation
through the walls of the accretion column can therefore be modeled
approximately using a term of the form
\begin{equation}
{\partial f \over \partial t} \Bigg|_{\rm approx}
= - \, {f \over t_{\rm esc}}
\ ,
\label{eq5.25}
\end{equation}
which is incorporated into the transport equation~(\ref{eq3.1}).

From a technical point of view, the utilization of the
escape-probability formalism implies that the distribution of the photon
number density across the column is proportional to the first
eigenfunction of the operator describing the diffusion of radiation
perpendicular to the column axis (e.g., Sunyaev \& Titarchuk 1980). In a
comprehensive, multidimensional calculation, the diffusion of photons
perpendicular to the column axis is treated by replacing the
escape-probability term $\partial f/\partial t = -f/t_{\rm esc}$ in the
transport equation~(\ref{eq3.1}) with the more rigorous expression
\begin{equation}
{\partial f \over \partial t} \Bigg|_{\rm exact}
= {1 \over r} \, {\partial\over\partial r}
\left({c \, r \over 3 \, n_e \, \sigma_\perp}\,{\partial f\over\partial r}
\right)
\ ,
\label{eq5.26}
\end{equation}
where $r$ is the radial coordinate measured from the central axis of the
cylindrical accretion column. If the gas density is constant across the
column at a given height, as assumed here, then equation~(\ref{eq5.26})
can be separated to obtain for the first eigenfunction the solution
\begin{equation}
f(r,t) = A \, \exp\left(- {c \, \xi^2 \, t \over 3 \, \ell_\perp}
\right) \, J_0\left({\xi \, r \over \ell_\perp}\right)
\ ,
\label{eq5.27}
\end{equation}
where $A$ is a constant, $\xi$ denotes the first eigenvalue, $J_0$
represents the zeroth-order Bessel function, and
\begin{equation}
\ell_\perp \equiv (n_e \, \sigperp)^{-1}
\label{eq5.28}
\end{equation}
is the electron scattering mean free path for photons propagating
perpendicular to the column axis.

The value of $\xi$ is determined by satisfying the flux boundary
condition at the surface of the column. By employing the Eddington
approximation, we can write the boundary condition as
\begin{equation}
\ell_\perp \, {\partial f \over \partial r}
+ \gamma \, f \, \bigg|_{r=r_0} = 0
\ ,
\label{eq5.29}
\end{equation}
where $\gamma$ is a positive constant of order unity and we have assumed
that no radiation is incident on the column from the outside. The precise value
of $\gamma$ is somewhat arbitrary, with Rybicki \& Lightman (1979)
setting $\gamma=3^{1/2}$ and Sunyaev \& Titarchuk (1980) setting $\gamma
= 3/2$. Combining equations~(\ref{eq5.27}) and (\ref{eq5.29}), we find
that $\xi$ is the smallest positive root of the equation
\begin{equation}
J_0\left({\xi \, r_0 \over \ell_\perp}\right)
= \left({\xi \over \gamma}\right)
J_1\left({\xi \, r_0 \over \ell_\perp}\right)
\ ,
\label{eq5.30}
\end{equation}
where $J_1$ denotes the order-unity Bessel function. 

The rate of change of the first eigenfunction (eq.~[\ref{eq5.27}]) is
given by
\begin{equation}
{\partial f \over \partial t}
= - {c \, \xi^2 \over 3 \, \ell_\perp} \, f(r,t)
\ .
\label{eq5.31}
\end{equation}
Comparison of equations~(\ref{eq5.25}) and (\ref{eq5.31}) reveals that
the escape-probability approximation and the rigorous diffusion model yield
the same result for the logarithmic time derivative $\partial \ln
f/\partial t$ if the mean escape time $t_{\rm esc}$ and the first
eigenvalue $\xi$ are related via
\begin{equation}
t_{\rm esc} = {3 \, \ell_\perp \over c \, \xi^2}
\ .
\label{eq5.32}
\end{equation}
This equation is well satisfied in all of our sample calculations, which
suggests that the escape-probability formalism provides a reasonable
description of the leakage of photons from the accretion column.

\section{CONCLUSIONS}

We have shown that the simplified model considered here, comprising a
radiation-dominated accretion column with a blackbody source/sink at its
base and a radiative shock, is able to reproduce the power-law continuum
spectra observed in accretion-powered X-ray pulsars with a range of
luminosities, as indicated in Figures~9 and 10. Our results represent
the first ab initio calculations of the X-ray spectrum associated with
the physical accretion scenario first suggested by Davidson (1973). For
given values of the stellar mass $M_*$ and the stellar radius $R_*$, our
model has only three free parameters, namely the column radius $r_0$,
the temperature at the top of the thermal mound $T_0$, and the accretion
rate $\dot M$. We have confirmed that the solution obtained for the
radiation spectrum is consistent with the dynamics of the flow in the
accretion column (see Fig.~11). Many pulsars are found to have spectra
that are well fitted using a combination of a power-law plus a blackbody
component, but until now this form has been adopted in a purely ad
hoc manner. The new model described here finally provides a firm
theoretical foundation for this empirical result. Our analytical
solution, based on a rigorous eigenfunction expansion, is relatively
straightforward to implement computationally because it avoids numerical
integration of the transport equation. The work presented here therefore
represents a significant step in the development of a comprehensive
theory for the spectral formation process in X-ray pulsars.

The photon spectral index $\alpha_0$ is related to the first eigenvalue
$\lambda_0$ via $\alpha_0 = \lambda_0 - 2$, and therefore the leading
eigenvalue determines the slope of the high-energy portion of the X-ray
spectrum. The value of $\alpha_0$ is quite sensitive to the strength of
the absorption at the thermal mound, which is determined by the
parameter $\beta$. According to equation~(\ref{eq5.13}), $\beta$ in turn
is a strong function of the mound temperature $T_0$. The higher the
temperature, the stronger the emission and absorption at the blackbody
surface of the mound. Model~2 has a higher temperature than model~1, and
therefore the value of $\beta$ is larger. Hence the spectrum is steeper
in model~2 because the photons spend less time on average upscattering
in the flow before either escaping through the column walls or being
``recycled'' by absorption at the mound location. Acceptable values for
$\alpha_0$, in the range between 2 and 4, are obtained for reasonable
source parameters, as indicated in Figures~2 and 6. Although a broad
range of spectral indices can be produced, we must have $\alpha_0 > 2$
in order to avoid an infinite photon energy density since the model
considered here does not include a high-energy cutoff.

The shape of the Green's function spectrum derived here is qualitatively
similar to the results obtained using the ``bulk motion Comptonization''
model coded into the XSPEC package using the BMC function (see Shrader
\& Titarchuk 1998, 1999; Borozdin et al. 1999). The BMC model includes
parameters corresponding to the high-energy spectral index, the
temperature of the soft photon source, the fraction of the Comptonized
component in the resulting spectrum, and the size of the emitting
region. However, the BMC model is designed to treat spectral formation
in black-hole accretion flows, whereas our focus here is on neutron star
accretion. Hence the model developed in this paper can be viewed as a
parallel formalism that is optimized for the treatment of the spectral
formation process in accretion-powered X-ray pulsars. In this sense, our
theory translates the free parameters of a generalized bulk
Comptonization model into specific values for the accretion rate,
temperature, and column radius for a particular X-ray pulsar.

Our goal in this paper is to explore the direct role of the accretion
shock in producing the observed continuum spectra in X-ray pulsars via
the first-order Fermi process. Although the model produces spectra that
are quite similar to those observed from several X-ray pulsars, it
cannot explain the spectra from other sources. For example, the observed
spectra for several of the brightest pulsars with luminosities in the
range $10^{37-38}\,{\rm ergs\,s}^{-1}$ (e.g., Her~X-1) are very flat
power-laws leading up to a sharp exponential cutoff. The photon index in
these sources is typically $\alpha_0 \sim 1$, which is outside the range
allowed by our model since it does not include a high-energy cutoff. In
order to understand spectral formation in the brightest pulsars,
additional effects such as thermal Comptonization, cyclotron features,
bremsstrahlung, and iron emission must be incorporated into the model.
In principle, one should also consider higher-order relativistic effects
that can alter the form of the fundamental transport equation (Psaltis \&
Lamb 1997). Thermal Comptonization is particularly important since this
process redistributes energy from the highest frequency photons to lower
energies via electron recoil. This naturally leads to the development of
a high-energy exponential turnover and a concomitant flattening of the
spectrum at moderate energies, as observed.

The thermal Comptonization process was studied by Lyubarskii \& Sunyaev
(1982) and Becker (1988) in the context of plane-parallel shocks.
However, the velocity profiles employed by these authors are not
consistent with the structure of an X-ray pulsar accretion column.
Poutanen \& Gierli\'nski (2003) computed X-ray pulsar spectra based on
the thermal Comptonization of soft radiation in a hot layer above the
magnetic pole, but their model did not include a complete treatment of
the bulk process, which is of crucial importance in X-ray pulsars. In
this paper we have carefully analyzed the effect of bulk Comptonization
on the emitted spectrum, while neglecting the corresponding thermal
process. The inclusion of thermal Comptonization is beyond the scope of
the present paper because it renders the transport equation insoluble
via analytical means. However, in future work we intend to develop a
comprehensive numerical model that incorporates both the bulk and
thermal processes.

The authors would like to thank Drs. Lev Titarchuk, Philippe Laurent,
and Nikos Kylafis who provided useful comments on the manuscript.
The authors are also grateful to the anonymous referee whose careful
reading and stimulating suggestions led to significant improvements
in the paper. PAB would also like to acknowledge generous support
provided by the Office of Naval Research.

\newpage

\section*{APPENDICES}

In the following sections we provide further details regarding the
various mathematical methods that form the foundation of our model.

\appendix

\section{Fundamental Solution in the Downstream Region}

When $y \ne y_0$, the spatial separation functions $g(\lambda,y)$
satisfy the homogeneous differential equation (cf. eq.~[\ref{eq4.12}])
\begin{equation}
y \, (1-y) \, {d^2 g \over dy^2}
+ \left({1 - 5 \, y \over 4}\right) {d g \over dy}
+ \left({\lambda \, y + y - 1 \over 4 \, y}\right) g
= 0
\ ,
\label{eqA.1}
\end{equation}
which has the fundamental solutions
\begin{equation}
\varphi_1(\lambda,y) \equiv y \, F(a, \, b; \, c; \, y)
\ ,
\label{eqA.2}
\end{equation}
\begin{equation}
\varphi^*_1(\lambda,y) \equiv y^{-1/4} \, F(a-5/4, \, b-5/4;
\, 2-c; \, y)
\ ,
\label{eqA.3}
\end{equation}
where $F(a,b;c;z)$ denotes the hypergeometric function (Abramowitz
\& Stegun 1970), and we have made the definitions
\begin{equation}
a \equiv {9 - \sqrt{17 + 16 \, \lambda} \over 8}\ , \ \ \ \ \ 
b \equiv {9 + \sqrt{17 + 16 \, \lambda} \over 8}\ , \ \ \ \ \ 
c \equiv {9 \over 4}
\ .
\label{eqA.4}
\end{equation}
As explained in the discussion following equation~(\ref{eq4.18}), in the
downstream limit the radiation pressure approaches the finite stagnation
value given by equation~(\ref{eq2.28}). The spatial separation functions
$g(\lambda,y)$ should mimic this behavior, and therefore we require that
$g$ approaches a constant in the limit $y \to 1$. This does not happen
in general, but only if the separation constant $\lambda$ is equal to one
of the eigenvalues $\lambda_n$, in which case we obtain the eigenfunction
\begin{equation}
g_n(y) \equiv g(\lambda_n,y)
\ .
\label{eqA.5}
\end{equation}
The hypergeometric functions appearing in equations~(\ref{eqA.2}) and
(\ref{eqA.3}) can be evaluated at $y=1$ using equation~(15.1.20) from
Abramowitz \& Stegun (1970), which gives for general values of $a$, $b$,
and $c$
\begin{equation}
F(a,b;c;1) = {\Gamma(c) \, \Gamma(c-a-b) \over \Gamma(c-a) \, \Gamma(c-b)}
\ .
\label{eqA.6}
\end{equation}
However, for the values of $a$, $b$, and $c$ in the current application,
we find that
\begin{equation}
c - a - b = 0
\ ,
\label{eqA.7}
\end{equation}
and therefore the hypergeometric functions $F(a, \, b; \, c; \, y)$ and
$F(a-5/4, \, b-5/4;\, 2-c; \, y)$ each {\it diverge} in the downstream limit
$y \to 1$.

Based on the asymptotic behaviors of the hypergeometric functions
appearing in the expressions for $\varphi_1$ and $\varphi^*_1$, we
conclude that in the downstream region ($y \ge y_0$), the eigenfunction
$g_n$ must be represented by a suitable linear combination of $\varphi_1$
and $\varphi^*_1$ that remains finite as $y \to 1$. In order to make further
progress, we need to employ equation~(15.3.10) from Abramowitz \& Stegun
(1970), which yields for general $a$, $b$, and $y$
\begin{eqnarray}
F(a, b\,; a+b\,; y) = {\Gamma(a+b) \over \Gamma(a) \, \Gamma(b)}
\ \sum_{n=0}^\infty \ {(a)_n \, (b)_n \over (n!)^2} \bigg[2 \Psi(n+1)
- \Psi(a+n)
\nonumber
\\
- \Psi(b+n) - \ln(1-y)\bigg] (1-y)^n
\ ,
\label{eqA.8}
\end{eqnarray}
where
\begin{equation}
\Psi(z) \equiv {1 \over \Gamma(z) } \, {d \Gamma(z) \over dz}
\ .
\label{eqA.9}
\end{equation}
By making use of this expression, we note that the logarithmic
divergences of the two functions $\varphi_1$ and $\varphi^*_1$ in the
limit $y \to 1$ can be balanced by creating the new function
\begin{equation}
\varphi_2(\lambda,y) \equiv {\Gamma(b) \over \Gamma(c) \, \Gamma(1-b)}
\ \varphi_1(\lambda,y)
- {\Gamma(1-a) \over \Gamma(2-c) \, \Gamma(a)}
\ \varphi^*_1(\lambda,y) \ ,
\label{eqA.10}
\end{equation}
which remains finite as $y \to 1$. Hence $\varphi_2$ represents the
fundamental solution for $g_n$ in the region downstream from the thermal
mound located at $y = y_0$. We can use the asymptotic behaviors of
$\varphi_1$ and $\varphi^*_1$ as $y \to 1$ to conclude that
\begin{equation}
\lim_{y \to 1} \ \varphi_2(\lambda,y) = {\pi \, [\cot(\pi \, a)
+ \cot(\pi \, b)] \over \Gamma(a) \, \Gamma(1-b)}
\ .
\label{eqA.11}
\end{equation}

Since the solutions $\varphi_1$ and $\varphi_2$ are applicable
in the upstream and downstream regions, respectively, the global
expression for the eigenfunction $g_n$ is therefore
\begin{equation}
g_n(y) = \cases{
\varphi_1(\lambda_n,y) \ , & $y \le y_0$ \ , \cr
B_n \, \varphi_2(\lambda_n,y) \ , & $y \ge y_0$ \ , \cr
}
\label{eqA.12}
\end{equation}
where the constant $B_n$ is evaluated using the continuity condition
\begin{equation}
B_n = {\varphi_1(\lambda_n,y_0) \over \varphi_2(\lambda_n,y_0)}
\ .
\label{eqA.13}
\end{equation}
It follows from equations~(\ref{eqA.11}), (\ref{eqA.12}), and
(\ref{eqA.13}) that the downstream value of $g_n$ is given by
\begin{equation}
\lim_{y \to 1} \ g_n(y)
= {\pi \, [\cot(\pi \, a)
+ \cot(\pi \, b)] \over \Gamma(a) \, \Gamma(1-b)}
\, {\varphi_1(\lambda_n,y_0) \over \varphi_2(\lambda_n,y_0)}
\ ,
\label{eqA.14}
\end{equation}
where $a$ and $b$ are evaluated using equations~(\ref{eq4.18}).

\section{Analytical Expression for the Wronskian}

The derivation of the eigenvalue equation~(\ref{eq4.24}) in \S~4.2 makes
use of the Wronskian of the two functions $\varphi_1$ and $\varphi_2$,
defined by
\begin{equation}
W(\lambda,y) \equiv
\varphi_1 \, {d\varphi_2 \over \partial y}
- \varphi_2 \, {d\varphi_1 \over \partial y}
\ .
\label{eqB.1}
\end{equation}
It is useful to derive an analytical expression for $W$ based on the
differential equation~(\ref{eq4.12}) governing the two functions
$\varphi_1$ and $\varphi_2$, which can be rewritten in the self-adjoint
form
\begin{equation}
{d \over dy}\left[y^{1/4} \, (1-y) \, {d\varphi \over dy}\right]
+ {\lambda \over 4 \, y^{3/4}} \, \varphi - T \, \varphi = 0
\ ,
\label{eqB.2}
\end{equation}
where
\begin{equation}
T \equiv {1-y \over 4 \, y^{7/4}} + {3 \, \beta \, v_0 \, \delta(y-y_0)
\over 7 v_c \, y^{3/4}}
\ .
\label{eqB.3}
\end{equation}
By applying equation~(\ref{eqB.2}) to the function $\varphi_2$ and
multiplying the result by $\varphi_1$, and then subtracting from
this the same expression with $\varphi_1$ and $\varphi_2$ interchanged,
we obtain
\begin{equation}
\varphi_1 \, {d \over dy}\left[y^{1/4} \, (1-y) \,
{d\varphi_2 \over dy}\right]
- \varphi_2 \, {d \over dy}\left[y^{1/4} \, (1-y) \,
{d\varphi_1 \over dy}\right] = 0
\ ,
\label{eqB.4}
\end{equation}
which can be rewritten as
\begin{equation}
y^{1/4} \, (1-y) \, {dW\over dy} + W \, {d\over dy}
\left[y^{1/4} \, (1-y)\right] = 0
\ ,
\label{eqB.5}
\end{equation}
where we have made use of the result
\begin{equation}
{dW \over dy} = \varphi_1 \, {d^2 \varphi_2 \over dy^2}
- \varphi_2 \, {d^2 \varphi_1 \over dy^2}
\ .
\label{eqB.6}
\end{equation}

Equation~(\ref{eqB.5}) can rearranged in the form
\begin{equation}
{d \ln W \over dy} = - {d \over dy} \ln\left[y^{1/4} \,
(1-y)\right]
\ ,
\label{eqB.7}
\end{equation}
which can be integrated to obtain the exact solution
\begin{equation}
W(\lambda,y) = {D(\lambda) \over y^{1/4} \, (1-y)}
\ ,
\label{eqB.8}
\end{equation}
where $D(\lambda)$ is an integration constant that may depend on
$\lambda$ but not on $y$. The exact dependence of $D$ on $\lambda$ can
be derived by analyzing the behaviors of the functions $\varphi_1$ and
$\varphi_2$ in the limit $y \to 0$. For small values of $y$, we have the
asymptotic expressions (Abramowitz \& Stegun 1970)
\begin{eqnarray}
\varphi_1 &\to& y \ , \phantom{SPAAAAAAAAAACE} y \to 0 \ ,
\nonumber
\\
\varphi_2 &\to& - \, {\Gamma(1-a) \over \Gamma(a) \,
\Gamma(2-c)} \ y^{-1/4} \ , \ \ \ \ \ y \to 0 \ ,
\label{eqB.9}
\end{eqnarray}
We therefore find that asymptotically,
\begin{equation}
W \to {5 \over 4} \, {\Gamma(1-a) \over \Gamma(a) \,
\Gamma(2-c)} \ y^{-1/4} \ , \ \ \ \ \ \ \ y \to 0
\ .
\label{eqB.10}
\end{equation}
Comparing this result with equation~(\ref{eqB.8}), we conclude that
\begin{equation}
D(\lambda) = {5 \over 4} \, {\Gamma(1-a) \over \Gamma(a) \,
\Gamma(2-c)}
\ ,
\label{eqB.11}
\end{equation}
and therefore the exact solution for the Wronskian is given by
\begin{equation}
W(\lambda,y) = {5 \over 4} \, {\Gamma(1-a) \over \Gamma(a) \,
\Gamma(2-c)} \ {y^{-1/4} \over 1-y}
\ .
\label{eqB.12}
\end{equation}
This result is used in \S~4.2 in the derivation of the eigenvalue
equation~(\ref{eq4.24}).

\section{Orthogonality of the Eigenfunctions}

The calculation of the expansion coefficients $C_n$ in the series representation
for the Green's function
\begin{equation}
\green(y_0,y,\epsilon_0,\epsilon)
= \epsilon^{-3} \sum_{n=0}^\infty \ C_n
\left(\epsilon \over\epsilon_0\right)^{3-\lambda_n} g_n(y)
\label{eqC.1}
\end{equation}
discussed in \S~4.3 depends on the establishment of the orthogonality of
the eigenfunctions $g_n(y)$. Here we shall demonstrate that the
eigenfunctions form an orthogonal set as required. Let us suppose that
$g_n(y)$ and $g_m(y)$ are two eigenfunctions corresponding to the
distinct eigenvalues $\lambda_n$ and $\lambda_m$, respectively. The
functions $g_n$ and $g_m$ satisfy the differential equation
equation~(\ref{eq4.12}), and therefore we can utilize the self-adjoint
form to write (cf. eq.~[\ref{eqB.2}])
\begin{equation}
g_m \left\{
{d \over dy}\left[y^{1/4} \, (1-y) \, {d g_n \over dy}\right]
+ {\lambda_n \over 4 \, y^{3/4}} \, g_n - T \, g_n
\right\} = 0
\ ,
\label{eqC.2}
\end{equation}
and
\begin{equation}
g_n \left\{
{d \over dy}\left[y^{1/4} \, (1-y) \, {d g_m \over dy}\right]
+ {\lambda_m \over 4 \, y^{3/4}} \, g_m - T \, g_m
\right\} = 0
\ ,
\label{eqC.3}
\end{equation}
where $T$ is given by equation~(\ref{eqB.3}). Subtracting the second equation
from the first yields, after integrating by parts with respect to $y$ from
$y=0$ to $y=1$,
\begin{equation}
(\lambda_n - \lambda_m) \int_0^1 y^{-3/4} \, g_n(y) \, g_m(y) \, dy
= 4 \, y^{1/4} \, (1-y) \left[
g_n \, {d g_m\over dy} - g_m \, {d g_n \over dy}
\right]\Bigg|_0^1
\ .
\label{eqC.4}
\end{equation}
Based on the asymptotic behaviors of the eigenfunctions $g_n$ and $g_m$
given by equations~(\ref{eqA.14}) and (\ref{eqB.9}), we find that the
right-hand side of equation~(\ref{eqC.4}) vanishes exactly, and we
therefore obtain
\begin{equation}
(\lambda_n - \lambda_m) \int_0^1 y^{-3/4} \, g_n(y) \, g_m(y) \, dy
= 0
\ ,
\label{eqC.5}
\end{equation}
which establishes the orthogonality of the eigenfunctions.

\section{Quadratic Normalization Integrals}

The evaluation of the Green's function $\green$ using equation~(\ref{eqC.1})
relies on knowledge of the expansion coefficients $C_n$, which are given by
\begin{equation}
C_n = {12 \, \dot N_0 \, y_0^{-3/4} \, g_n(y_0) \over
7 \, \pi \, r_0^2 \, v_c \, I_n}
\ ,
\label{eqD.1}
\end{equation}
where the quadratic normalization integrals $I_n$ are defined by
\begin{equation}
I_n \equiv \int_0^1 y^{-3/4} \, g_n^2(y) \, dy
\ .
\label{eqD.2}
\end{equation}
The direct computation of the normalization integrals $I_n$ via
numerical integration is costly and time consuming, and therefore it is
desirable to have an alternative procedure available for their
evaluation. In fact, it is possible to derive an analytical expression
for the normalization integrals based on manipulation of the fundamental
differential equation~(\ref{eq4.12}) governing the eigenfunctions
$g_n(y)$.

Let us suppose that $g(\lambda,y)$ is a general solution to
equation~(\ref{eq4.12}) for an arbitrary value of $\lambda$ (i.e., not
necessarily an eigenvalue) that has the asymptotic upstream behavior
\begin{equation}
g(\lambda,y) \to y \ , \ \ \ \ \ \ \ \ y \to 0
\ .
\label{eqD.3}
\end{equation}
We shall also require that $g$ be continuous at $y=y_0$, and that
it satisfy the derivative jump condition (see eq.~[\ref{eq4.14}])
\begin{equation}
\lim_{\varepsilon \to 0} \
{d g \over d y}\Bigg|_{y=y_0+\varepsilon}
- {d g \over d y}\Bigg|_{y=y_0-\varepsilon}
= {3 \, \beta \over 4 \, y_0} \ g(\lambda,y_0)
\ .
\label{eqD.4}
\end{equation}
After a bit of algebra, we find that the global solution for $g$ can
be expressed as
\begin{equation}
g(\lambda,y) = \cases{
\varphi_1(\lambda,y) \ , & $y \le y_0$ \ , \cr
(1+\hat a) \, \varphi_1(\lambda,y) + \hat b \, \varphi_2(\lambda,y)
\ , & $y \ge y_0$ \ , \cr
}
\label{eqD.5}
\end{equation}
where $\hat a$ and $\hat b$ are given by
\begin{equation}
\hat a = - \, {3 \, \beta \varphi_1(\lambda,y_0)
\, \varphi_2(\lambda,y_0)
\over 4 \, y_0 W(\lambda,y_0)} \ , \ \ \ \ \ \ \ 
\hat b = \ \ \ {3 \, \beta \varphi_1^2(\lambda,y_0) \over
4 \, y_0 W(\lambda,y_0)} \ ,
\label{eqD.6}
\end{equation}
and the Wronskian $W$ is evaluated using equation~(\ref{eqB.12}).

Comparing the general solution for $g(\lambda,y)$ with the solution
for the eigenfunction $g_n(y)$ given by equation~(\ref{eqA.12}), we
note that
\begin{equation}
\lim_{\lambda \to \lambda_n} \hat a = -1 \ , \ \ \ \ \ \ 
\lim_{\lambda \to \lambda_n} \hat b = B_n
\ .
\label{eqD.7}
\end{equation}
We can now use the self-adjoint form of equation~(\ref{eq4.12})
to write (cf. eqs.~[\ref{eqC.2}] and [\ref{eqC.3}])
\begin{equation}
g_n \left\{
{\partial \over \partial y}\left[y^{1/4} \, (1-y) \,
{\partial g \over \partial y}
\right]
+ {\lambda \over 4 \, y^{3/4}} \, g - T \, g
\right\} = 0
\ .
\label{eqD.8}
\end{equation}
and
\begin{equation}
g \left\{
{d \over dy}\left[y^{1/4} \, (1-y) \, {d g_n \over dy}\right]
+ {\lambda_n \over 4 \, y^{3/4}} \, g_n - T \, g_n
\right\} = 0
\ ,
\label{eqD.9}
\end{equation}
Subtracting the second equation from the first and integrating by parts
from $y=0$ to $y=1$ yields
\begin{equation}
(\lambda - \lambda_n) \int_0^1 y^{-3/4} \, g(\lambda,y)
\, g_n(y) \, dy
= 4 \, y^{1/4} \, (1-y) \left[
g(\lambda,y) \, {d g_n\over dy} - g_n(y) \, {\partial g \over \partial y}
\right]\Bigg|_0^1
\ ,
\label{eqD.10}
\end{equation}
Since $g \to y$ and $g_n \to y$ as $y \to 0$, we conclude that the evaluation
at the lower bound $y=0$ on the right-hand side yields zero, and consequently
in the limit $\lambda \to \lambda_n$ we obtain for the quadratic normalization
integral $I_n$ (see eq.~[\ref{eqD.2}])
\begin{equation}
I_n = \int_0^1 y^{-3/4} \, g_n^2(y) \, dy
= \lim_{\lambda \to \lambda_n}
{4 \, y^{1/4} \, (1-y) \left[
g(\lambda,y) \, (d g_n / dy) - g_n(y) \, (\partial g / \partial y)
\right] \over \lambda - \lambda_n}\Bigg|_{y=1}
\ .
\label{eqD.11}
\end{equation}

The numerator and denominator on the right-hand side of
equation~(\ref{eqD.11}) each vanish in the limit $\lambda \to
\lambda_n$, and therefore we can employ L'H\^opital's rule to show that
(e.g., Becker 1997)
\begin{equation}
I_n = \lim_{\lambda \to \lambda_n}
4 \, y^{1/4} \, (1-y) \left[
{\partial g \over \partial y} \, {d g_n \over dy} - g_n \,
{\partial^2 g \over \partial y \, \partial \lambda}
\right]\Bigg|_{y=1}
\ .
\label{eqD.12}
\end{equation}
Substituting the analytical forms for $g_n(y)$ and $g(\lambda,y)$ given
by equations~(\ref{eqA.12}) and (\ref{eqD.5}), respectively, we find
that equation~(\ref{eqD.12}) can be rewritten as
\begin{equation}
I_n = \lim_{y \to 1} \
4 \, y^{1/4} \, (1-y) \, B_n \left[
W(\lambda,y) \, {d \hat a \over d \lambda}
+ B_n \, {\partial\varphi_2 \over \partial \lambda} \,
{\partial\varphi_2 \over \partial y}
- B_n \, \varphi_2(\lambda,y) \, {\partial^2 \varphi_2 \over
\partial y \, \partial \lambda}\right]
\Bigg|_{\lambda=\lambda_n}
\ ,
\label{eqD.13}
\end{equation}
where we have also utilized equations~(\ref{eqB.1}) and (\ref{eqD.6}).
Based on the asymptotic behavior of $\varphi_2$ given by
equation~(\ref{eqA.11}), we conclude that the final two terms on the
right-hand side of equation~(\ref{eqD.13}) contribute nothing in the
limit $y \to 1$, and therefore our expression for $I_n$ reduces to
\begin{equation}
I_n = \lim_{y \to 1} \
4 \, y^{1/4} \, (1-y) \, B_n \, W(\lambda,y)
{d \hat a \over d \lambda}
\Bigg|_{\lambda=\lambda_n}
\ .
\label{eqD.14}
\end{equation}

Since $y = 1$ is a singular point of the differential equation
(\ref{eq4.12}), it is convenient to employ the relation
(see eq.~[\ref{eqB.8}])
\begin{equation}
W(\lambda,y) \ y^{1/4} \, (1-y)
= W(\lambda,y_0) \ y_0^{1/4} \, (1-y_0)
\ ,
\label{eqD.15}
\end{equation}
which allows us to transform the evaluation in equation~(\ref{eqD.14})
from $y=1$ to $y=y_0$ to obtain the equivalent result
\begin{equation}
I_n = 4 \, y_0^{1/4} \, (1-y_0) \, \hat a \, W(\lambda,y_0) \,
{\varphi_1(\lambda_n,y_0)
\over \varphi_2(\lambda_n,y_0)} \,
{d \ln \hat a \over d \lambda}
\Bigg|_{\lambda=\lambda_n}
\ ,
\label{eqD.16}
\end{equation}
where we have also substituted for $B_n$ using equation~(\ref{eqA.13}).
The derivative on the right-hand side can be evaluated using
equation~(\ref{eqD.6}), which yields
\begin{equation}
{d \ln \hat a \over d \lambda}
= {\partial\ln\varphi_1 \over \partial\lambda}
+ {\partial\ln\varphi_2 \over \partial\lambda}
- {\partial \ln W \over \partial \lambda}
\ ,
\label{eqD.17}
\end{equation}
where the derivative of the Wronskian is given by (see eqs.~[\ref{eqA.4}]
and [\ref{eqB.12}])
\begin{equation}
{\partial \ln W \over \partial \lambda}
= {\Psi(a) + \Psi(1-a) \over (17 + 16 \, \lambda)^{1/2}}
\ ,
\label{eqD.18}
\end{equation}
and
\begin{equation}
\Psi(z) \equiv {1 \over \Gamma(z)} \, {d \Gamma(z) \over dz}
\ .
\label{eqD.19}
\end{equation}
Combining equations~(\ref{eqD.6}), (\ref{eqD.16}), (\ref{eqD.17}), and
(\ref{eqD.18}), we find that that the quadratic normalization integrals
can be evaluated using the closed-form expression
\begin{equation}
I_n = K(\lambda_n,y_0)
\ ,
\label{eqD.20}
\end{equation}
where
\begin{equation}
K(\lambda,y) \equiv
3 \, \beta \, y^{-3/4} (1-y) \, \varphi_1^2(\lambda,y)
\left[{\Psi(a) + \Psi(1-a) \over (17 + 16 \, \lambda)^{1/2}}
- {\partial\ln\varphi_1\over\partial\lambda}
- {\partial\ln\varphi_2\over\partial\lambda}
\right]
\ .
\label{eqD.21}
\end{equation}
This formula provides an extremely efficient alternative to numerical
integration for the computation of $I_n$.

\section{Column-Integrated Green's Function}

In \S~4.5 we demonstrated that the column-integrated Green's function
$\greencolumn(y_0,\epsilon_0,\epsilon)$ can be
written as (see eq.~[\ref{eq4.38}])
\begin{equation}
\greencolumn(y_0,\epsilon_0,\epsilon)
= {\dot N_0 \over \epsilon \, y_0^{3/4}}
\ \sum_{n=0}^\infty \ {g_n(y_0) \over I_n}
\left(\epsilon \over\epsilon_0\right)^{3-\lambda_n}
\, X_n
\ ,
\label{eqE.1}
\end{equation}
where the quadratic normalization integrals $I_n$ are computed using
equation~(\ref{eq4.29}), and
\begin{equation}
X_n \equiv \int_0^1 g_n(y) \, (1-y) \, y^{-1} \, dy
\ .
\label{eqE.2}
\end{equation}
As an alternative to numerical integration for the evaluation of the
integral $X_n$, here we shall derive an analytical expression for this
quantity. First we use equation~(\ref{eq4.20}) to substitute for $g_n(y)$
in equation~(\ref{eqE.2}) to obtain
\begin{equation}
X_n = \int_0^{y_0} \varphi_1(\lambda_n,y) \, (1-y) \, {dy \over y}
+ {\varphi_1(\lambda_n,y_0) \over \varphi_2(\lambda_n,y_0)}
\int_{y_0}^1 \varphi_2(\lambda_n,y) \, (1-y) \, {dy \over y}
\ .
\label{eqE.3}
\end{equation}
Substitution for $\varphi_2$ using equation~(\ref{eq4.19}) and rearranging
terms, we can write
\begin{eqnarray}
X_n &=& L_1 \, {\varphi_1(\lambda_n,y_0) \over \varphi_2(\lambda_n,y_0)}
\int_0^1 \varphi_1(\lambda_n,y) \, (1-y) \, {dy \over y}
- L_2 {\varphi_1^*(\lambda_n,y_0) \over \varphi_2(\lambda_n,y_0)}
\int_0^{y_0} \varphi_1(\lambda_n,y) \, (1-y) \, {dy \over y}
\nonumber
\\
&-& L_2 {\varphi_1(\lambda_n,y_0) \over \varphi_2(\lambda_n,y_0)}
\int_{y_0}^1 \varphi_1^*(\lambda_n,y) \, (1-y) \, {dy \over y}
\ ,
\label{eqE.4}
\end{eqnarray}
where
\begin{equation}
L_1 \equiv {\Gamma(b) \over \Gamma(c) \, \Gamma(1-b)} \ , \ \ \ \ \ \ \ 
L_2 \equiv {\Gamma(1-a) \over \Gamma(2-c) \, \Gamma(a)}
\ .
\label{eqE.5}
\end{equation}
To make further progress, we need to evaluate the fundamental indefinite
integrals
\begin{equation}
K_1 \equiv \int \varphi_1(\lambda_n,y) \, (1-y) \, {dy \over y} \ ,
\ \ \ \ \ \ \
K_2 \equiv \int \varphi_1^*(\lambda_n,y) \, (1-y) \, {dy \over y}
\ .
\label{eqE.6}
\end{equation}
Fortunately, these integrals can be carried out analytically. By applying
equation~(15.2.1) from Abramowitz \& Stegun (1970) twice, we obtain for
$K_1$ the result
\begin{eqnarray}
K_1 = {(c-1) \over (a-1) \, (b-1)}
\, {(c-2) \over (a-2) \, (b-2)}
\, F(a-2, b-2; c-2; y)
\nonumber
\\
+ {(1-y) \, (c-1) \over (a-1) \, (b-1)}
\, F(a-1, b-1; c-1; y)
+ {\cal C}_1
\ ,
\label{eqE.7}
\end{eqnarray}
where ${\cal C}_1$ is an integration constant. Likewise, we can apply
equation~(15.2.4) from Abramowitz \& Stegun (1970) twice to find that
\begin{eqnarray}
K_2 &=& {(1-y) \, y^{2-c} \over 2-c}
\, F(1-a, 1-b\,; 3-c\,; y)
\nonumber
\\
&+& {y^{3-c} \over (2-c) \, (3-c)} \,
F(1-a\,, 1-b\,; 4-c\,; y)
+ {\cal C}_2
\ ,
\label{eqE.8}
\end{eqnarray}
where ${\cal C}_2$ is another integration constant. Combining relations
and simplifying, after some algebra we obtain the final result
\begin{eqnarray}
X_n &=& {\Gamma(1-a) \, (1-y_0)^2 \over \Gamma(a) \,
\varphi_2(\lambda_n,y_0)}
\bigg\{{y_0^{2-c} \, \varphi_1(\lambda_n,y_0) \over \Gamma(3-c)}
\bigg[
F(2-a,2-b;3-c;y_0)
\nonumber
\\
&+& {y_0 \over 3-c} \, F(3-a,3-b;4-c;y_0) \bigg]
+ {\varphi_1^*(\lambda_n,y_0) \over (a-1) \, (b-1) \, \Gamma(1-c)}
\bigg[
F(a,b\,;c-1;y_0)
\nonumber
\\
&+& {(c-2) \over (a-2) \, (b-2)} \, F(a,b;c-2;y_0)\bigg]
\bigg\}
+ {(1-c) \over (a-1) \, (b-1)}
\left[1 + {(c-2) \over (a-2) \, (b-2)}\right]
\ .
\qquad
\label{eqE.9}
\end{eqnarray}
This formula allows the efficient computation of the integrals $X_n$
appearing in equation~(\ref{eq4.38}) for the column-integrated Green's
function.


\clearpage

\begin{deluxetable}{cccccccccccccc}
\tabletypesize{\scriptsize}
\tablecaption{Model Parameters\label{tbl-1}}
\tablewidth{0pt}
\tablehead{
\colhead{Model}
& \colhead{$r_0$ (km)}
& \colhead{$T_0$ (K)}
& \colhead{$\dot M$ (g\,s$^{-1}$)}
& \colhead{$p$}
& \colhead{$q$}
& \colhead{$y_0$}
& \colhead{$\beta$}
& \colhead{$\alpha_0$}
& \colhead{$h_0/x_{\rm st}$}
& \colhead{$v_0/c$}
}
\startdata
1
&6.0
&$7.3 \times 10^6\,$
&$2.69 \times 10^{16}$
&0.890
&1.087
&0.999770
&$2.645 \times 10^1$
&2.0398
&$2.71 \times 10^{-4}$
&$1.48 \times 10^{-4}$
\\
2
&1.3
&$9.0 \times 10^6\,$
&$3.23 \times 10^{13}$
&0.286
&0.954
&0.999998
&$2.894 \times 10^5$
&2.6382
&$2.36 \times 10^{-6}$
&$1.22 \times 10^{-6}$
\\
\enddata



\end{deluxetable}

\end{document}